\documentclass[a4paper,11pt]{article}
\usepackage{jcappub} 

\title{MARVELously Dark: the density profile evolution of dwarf halos in velocity-dependent SIDM}

\usepackage{tikz,xcolor,hyperref}
\usepackage{soul}

\definecolor{lime}{HTML}{A6CE39}
\DeclareRobustCommand{\orcidicon}{
	\begin{tikzpicture}
	\draw[lime, fill=lime] (0,0) 
	circle [radius=0.16] 
	node[white] {{\fontfamily{qag}\selectfont \tiny ID}};
	\draw[white, fill=white] (-0.0625,0.095) 
	circle [radius=0.007];
	\end{tikzpicture}
	\hspace{-2mm}
}

\foreach \x in {A, ..., Z}{\expandafter\xdef\csname orcid\x\endcsname{\noexpand\href{https://orcid.org/\csname orcidauthor\x\endcsname}
			{\noexpand\orcidicon}}
}

\author[a,1]{{Anna Engelhardt\orcidA{}} \note{Corresponding author.}}
\author[a]{Ferah Munshi\orcidF{}}
\author[b,c]{Annika H. G. Peter\orcidD{}}
\author[d]{Ethan~O.~Nadler\orcidG{}} 
\author[e,f]{Akaxia Cruz\orcidB{}}
\author[g]{Alyson M. Brooks\orcidC{}}
\author[h]{Zhichao Carton Zeng\orcidH{}}
\author[i]{Thomas R. Quinn\orcidE{}}
\author[a]{Blake Keith\orcidI{}}

\affiliation[a]{Department of Physics \& Astronomy, George Mason University, \\4400 University Drive, MSN: 3F3 Fairfax, VA 22030-4444, USA}
\affiliation[b]{Department of Physics and Center for Cosmology and AstroParticle Physics, The Ohio State University, \\191 W. Woodruff Ave., Columbus, OH 43210}
\affiliation[c]{Department of Astronomy, The Ohio State University, \\140 18th Ave. W., Columbus, OH 43210}
\affiliation[d]{Department of Astronomy \& Astrophysics, University of California, \\San Diego, La Jolla, CA 92093, USA} 
\affiliation[e]{Center for Computational Astrophysics, Flatiron Institute, \\162 Fifth Ave, New York, NY 10010, USA}
\affiliation[f]{Department of Physics, Princeton University, \\Princeton, NJ 08544, USA}
\affiliation[g]{Department of Physics \& Astronomy, Rutgers, The State University of New Jersey, \\136 Frelinghuysen Road, Piscataway, NJ 08854, USA}
\affiliation[h]{Department of Physics \& Astronomy, Texas A\&M University, \\400 Bizzell St. College Station, TX. 77843, USA}
\affiliation[i]{Astronomy Department, University of Washington, \\Box 351580, Seattle, WA, 98195-1580}

\emailAdd{aengelha@gmu.edu}

\abstract{Self-interacting dark matter (SIDM) with a sufficiently large cross section has been shown to naturally produce constant dark matter (DM) cores, as well as core-collapse, at the centers of dwarf halos on cosmic timescales, potentially reducing tensions with observation. Here, we present halos from a new dark matter only (DMO) cosmological (SIDM) simulation: Ms.Marvel DMO with a velocity-dependent self-interaction cross section with $\sigma/m_\text{max} = 50$ cm$^2$/g at $v_\text{max} = 35$ km/s. We compare these to the CDM suite of Storm simulations including both DMO and dark matter + hydrodynamics runs, in order to test core-formation (and core-collapse) across different dark matter models. We show that Ms.Marvel DMO can reproduce core slopes consistent with observations of isolated dwarf galaxies and more massive ($\text{M}_{vir} \gtrsim 10^{10} M_{\odot}$) CDM dwarf halos that include stellar feedback from the matched CDM run (Storm CDM+baryons). We identify nine Ms.Marvel SIDM DMO halos in the core-collapse phase of gravothermal evolution with halo masses below $2\times 10^9 M_{\odot}$. We find that using the inner density slope to measure the core-collapse timescales of Ms.Marvel DMO halos agrees well with predicted collapse times estimated with the parametric model for SIDM halos introduced by \cite{Yang2023}. Additionally, compared to central density, the inner density slope is less sensitive to both the radius of measurement and halo merger history. These results indicate that the slope of the inner DM density profile more cleanly differentiates core-collapsed versus core-forming halos than central density amplitude.}

\begin{document}

\maketitle
\flushbottom

\section{Introduction}
\label{sec:intro}
Dwarf galaxies, being the most numerous and dark matter (DM)-dominated systems in the universe, have long been a fertile testing ground for the cold dark matter (CDM) paradigm. Although the CDM framework has been largely successful in explaining the observed universe \citep[see][]{Planck16}, several issues arose when observations and simulations began to thoroughly explore the small-scale regime. Contentions such as the ``missing satellites'' problem \citep{Kauffmann93,Klypin99,Moore99}, the ``too big to fail'' problem \citep{BoylanKolchin11,Papastergis15}, and the ``cusp-core'' problem \citep{Moore94,Burkert95,Bullock17} all call into question the fidelity of both cosmological simulations and the CDM framework itself. These tensions are primarily alleviated in the CDM paradigm through the inclusion of baryonic physics, with a focus on stellar feedback, in cosmological simulations \citep{White91,Navarro96,Gelato1999,read05,Quinn96,Somerville08,Governato10,Oh2011,Hopkins11,Pontzen12,Zolotov12,Brooks13,brooks2014,Brook_2015,Chan2015,papastergis2016,sawala2016,Wetzel16,Brooks17,GarrisonKimmel19,Munshi21}. However, alternative DM models, such as self-interacting dark matter (SIDM), remain valid possibilities \citep{Spergel00,Dave01,Colin2002,Ahn05,Rocha2013,Peter13,kaplinghat2014,Kaplinghat2016}. Over the past decade, particle physicists have begun studying a broader range of DM models, many of which naturally produce self-interactions \citep{Feng_2009,Buckley_2010,loeb2011,Zavala2013,Kaplinghat_2014_direct,Kouvaris_2015,Bernal_2016,Kainulainen_2017,Bringmann_2017,Cirelli_2017,Kahlhoefer_2017,Hambye_2020,Dutta_2021,Borah_2022,garcíacely2025piondarkmattertheta,dev2025newconstraintsneutrinodarkmatter}. Thus, the in-depth study of dwarf galaxies simulated with alternative DM is complementary to the efforts being made by particle physicists to constrain DM models. 
\newline\indent Previous work \citep{Bursting2024,Mostow_2024_bursts} shows that, in CDM halos, cores are formed and sustained through the stellar feedback induced by repeated bursts of star formation. As a result, in low-mass halos, which have fewer stars and less extended star formation histories (SFHs), baryonic feedback is insufficient to convert cusps into cores. However, \cite{Almeida2024dens} and \cite{Almeida2024bright} find evidence of central density cores in observed dwarf satellite galaxies, with stellar masses of $10^{3-4}M_\odot$ and $10^{6-8}M_\odot$ respectively, from their surface brightness profiles and stellar surface density profiles. For low-luminosity galaxies with $M_B \gtrsim -19$, \cite{de_Block_high_2008} find that dark matter halos prefer cored profiles over cusped profiles. Additionally, \cite{Oh2015} find evidence of central DM density slopes, the slope of the DM density profile at inner radii, that are shallower than predictions from simulations \citep{DiCintio14,Lazar2020} between stellar masses of $10^{5-8}M_\odot$. The rotation curves of observed spiral galaxies may also be more diverse than predictions from CDM simulations \citep{Oman_2015,Ren_2019,Dehghani_2020,Santos_santos_2020}, however, \cite{cruz2025dwarfdiversitylambdacdmbaryons} find that dwarf galaxies from the MARVELous dwarf simulations are capable of reproducing the observed diversity of rotation curves in the $\Lambda$CDM paradigm. 
\newline\indent These tensions with the $\Lambda$CDM paradigm may be resolved by alternative DM models such as SIDM. In SIDM halos, self-scattering among the DM particles causes the central halo to dynamically heat, leading to heat inflow towards the center of the halo. The subsequent negative heat capacity at the center of the halo causes a runaway gravothermal collapse process, with core-collapsed halos becoming denser and cuspier in the central region than their CDM counterparts \citep{Lynden-Bell1968,Kochanek2000,Balberg2002,Koda2011,Pollack2015,Essig2019,Zavala2019,Nishikawa2020,Sameie2020,Correa2021,Turner2021,Carton2022,Tran2024,Mace2024,Palubski2024,Fischer_2024,Tran2025}. SIDM halos can successfully reproduce the observed diversity of galactic rotation curves \citep{Zentner2022,Ren_2019,Kaplinghat2020,Shen2024,Roberts_2025,Nadler2025}. Also, \cite{Oh2011} find that observed galaxies are better fit by an isothermal core, such as those that are naturally formed in SIDM halos. SIDM simulations successfully reproduce the cored DM density profiles of observed bright dwarfs
\citep{Burkert2000,Spergel00,Colin2002,Bellazzini2013,Rocha2013,Fry2015,Elbert2015,Kaplinghat2016,Tulin18,Robertson2021,Shen2024}, and reproduce the successes of CDM at galaxy cluster mass scales \citep{Peter13,Zavala2013}. Also, \cite{Oh2011} find that observed galaxies are better fit by an isothermal core, such as those that are naturally formed in SIDM halos. 
SIDM simulations have also been shown to successfully match the observed diversity of DM profiles, subhalo mass function, and the radial number density of the Milky Ways (MWs) satellite population \citep{Vogelsberger2012,Correa2021,Rahimi2023,Yang_2023,Nadler2025,Nadler2024}. Furthermore, the Milky Way and its satellite population can be used to constrain the interacting cross section of SIDM \citep{Correa2021,Kim_MilkyWay_2021,Nadler20,ONeil2023,Slone2023, Nadler2025}. For example, \cite{Silverman_2023} find that for interacting cross sections $\sigma /m \lesssim 5$~cm$^2$g$^{-1}$ SIDM halos are unable to reproduce the densest ultra-faint and classical dwarf spheroidal galaxies in the MW.
\newline\indent New and upcoming surveys such as the Vera Rubin Observatory (Rubin) and {\it Roman Space Telescope} will enable observers to probe dark matter abundance and structure through strong gravitational lensing as well as improved dwarf galaxy and stellar stream measurements \citep{chakrabarti2022snowmass2021cosmicfrontierwhite}. The Legacy Survey of Space and Time (LSST) with Rubin will have high enough resolution to allow precise measurements of the density profiles and shapes of dark matter in dwarf galaxy halos with stellar masses down to $10^8 M_\odot$ based on weak lensing \citep{mao2022snowmass2021veracrubin,Leauthaud_Deep_2020} and discover many more dwarf galaxies~\citep{Mutlu_Pakdil_2021_resolved} and MW satellites \citep{Tsiane_2025}. Additionally, extremely large telescopes (ELTs) will allow observers to probe the dark matter-subhalo mass function down to halo masses of $10^6 M_\odot$ by targeting the MWs stellar streams and its Galactic acceleration field \citep{chakrabarti2022realtimecosmologyhighprecision}. Constraints on the halo mass function and precise measurements of the DM density profiles of the halos of dwarf galaxies can therefore be used to constrain DM models. 
\newline\indent Previous work constrains the interacting cross section for SIDM models by comparing predictions from simulations with observed galaxies. For small cross sections, SIDM halos replicate the successful predictions of large-scale structure in the CDM paradigm \citep{Rocha2013,Cyr_Racine_2016,adhikari2022astrophysicaltestsdarkmatter}, while large cross sections at low velocities naturally create cored and core-collapsed central density profiles at dwarf galaxy scales \citep{Burkert2000,Spergel00,Colin2002,Vogelsberger2012,Bellazzini2013,Rocha2013,Zavala2013,Fry2015,Elbert2015,Correa2021,Kaplinghat2016,Tulin18,Robertson2021,Shen2024}. At galaxy cluster mass scales, studies of the inner DM density profiles of galaxy clusters place strong constraints on the interaction cross section for SIDM ($\sigma /m \lesssim 0.1 -1.0 $~cm$^2$g$^{-1}$); \citep{Andrade2022,Eckert2022,Rocha2013,Elbert2015}. At smaller scales, SIDM has been shown to naturally produce constant density cores at the centers of isolated dwarf halos with cross sections ranging from $0.5-50$ cm$^2$g$^{-1}$, reducing tensions with the ``too big to fail" and ``cusp-core" problems \citep{Vogelsberger2012,Zavala2013,Fry2015,Elbert2015} for classical dwarfs. However, \cite{Kim_MilkyWay_2021} find that the centers of ultra-faints are cusped and rule out SIDM models with a constant cross section $\ge 0.1 \text{cm}^2\text{g}^{-1}$. These varying constraints on the cross section of interaction at different mass scales motivate the study of a velocity-dependent cross section of interaction. 
\newline\indent Velocity-dependent SIDM models with gravothermally induced core-collapse may also be able to account for the observed diversity of rotation curves of MW satellites \citep{Zavala2019,Correa2021,Turner2021, Yang2023}. Recent observations indicate that the the densities of both satellite dwarfs \citep{Zavala2019} and isolated halos \citep{Oman_2015,Santos_santos_2020} are varied. Velocity-dependent SIDM models with sufficiently large interacting cross sections at low relative velocities can induce gravothermal core-collapse in ultra-faint dwarfs while preserving core formation in classical mass halos. Core-collapse timescales vary by halo concentration and can be impacted by tidal effects \citep{Carton2022,Jiang19,Benavides2021,Benavides2023}. As a result, SIDM models with large cross sections that drive core collapse at low masses produce diverse DM density profiles ranging from cored to steeply cusped. 
\newline\indent Previous works use parametric models of gravothermal collapse \citep{Yang2023,Yang2024,Yang2025} or idealized simulations to predict core-collapse times of SIDM halos with various cross sections \citep{Mace2024,Carton2022,zeng2025tillcorecollapsesevolution}. Additionally, various velocity-dependent SIDM models have been studied in cosmological simulations of MW satellites \citep{Vogelsberger2012,Correa2021,Rahimi2023,Yang_2023,Nadler2025,Nadler2024}. However, this work compares core-collapse timescales from a statistical sample of isolated halos with varied merger histories from a zoom-in simulation and contextualize our results with predictions from analytic calculations. We present a sample of halos from Ms.Marvel DMO, an SIDM cosmological volume simulation with a velocity-dependent cross section and a Yukawa potential \citep{Feng2010,loeb2011} in the classical regime. We contextualize our findings using results from the Storm simulations, a suite of matching CDM simulations run with the same initial conditions as the Ms.Marvel box. These runs are a combination of dark-matter-only (DMO) simulations and simulations run with baryonic components.
\newline\indent The focus of this paper is to test core-collapse timescales of SIDM halos with a velocity-dependent interacting cross section. \cite{Yang2025} validate the parametric model in cosmological simulations of massive ($10^{12-13} M_\odot$) host halos and their surrounding environment, finding predictions to be within an agreement of $10\% - 50\%$ for most (sub)halos. Results from \cite{Nadler2024,Nadler2025} further test the accuracy of the parametric model by analytically calculating the gravothermal evolution of SIDM (sub)halos using matched CDM counterparts in cosmological zoom-in simulations of host halos with masses of $10^{11}, 10^{12}, 10^{13}$, and $10^{14} M_\odot$ and their satellite populations. They find that the predictions accurately model gravothermal evolution through core-collapse. This work will test analytic predictions from the parametric model against cosmological simulations of low-mass, dwarf halos in isolated field environments. 
\newline\indent The paper is outlined as follows; in Section \ref{sec:sim} we describe the simulations including our implementation of relevant baryonic physics and a velocity-dependent SIDM model. We show our results in Section \ref{sec:results}, provide a detailed discussion in Section \ref{sec:discussion}, and give a brief overview of our findings in Section \ref{sec:conclusions}. 

\section{Simulations and Methodology}
\label{sec:sim}
\indent This work studies the gravothermal evolution of SIDM halos with a velocity-dependent cross section of interaction from the zoom-in cosmological simulation Ms.Marvel DMO. We contextualize our findings with results from corresponding CDM simulations, Storm CDM DMO and Storm CDM+baryons, to serve as a benchmark comparison. These results focus on differences in core formation timescales and the inner slope of the DM density profiles of dwarf galaxies between the two DM models. 
\newline\indent The Ms.Marvel simulation has the same initial cosmological conditions as the Storm simulation from the `MARVEL-ous Dwarfs'' suite \citep[hereafter ``Marvel'';][]{Bellovary2019,Munshi2019,Munshi21,Christensen2023,Bursting2024,Piacitelli2025,Keith2025,Ruan2025}, but an alternative DM model. This suite consists of four cosmological zoom-in simulations with a WMAP3 cosmology \citep{Spergel07} where, rather than simulating individual galaxies, small volumes with multiple dwarf galaxies and maximum halo masses of approximately $10^{11} M_\odot$ are simulated at high resolution. The high resolution of the simulations coupled with the isolated environment make these ideal simulations for studying the evolution of dwarf galaxies in relative isolation. We exclude dwarf halos identified as satellites in our sample, minimizing the potential gravitational effects of host halos on our sample. However, the halos in our sample have varied merger histories, including major and minor mergers or potential fly-by interactions that may impact the morphology of halos. For additional details refer to Section \ref{subsec:trace}. The simulations are run with the Nbody+SPH code C{\sc ha}NG{\sc a} \citep{Changa} which utilizes the hydrodynamic modules from GASOLINE2 \citep{Wadsley04,Wadsley17}.

\subsection{Halo identification and sample}
\label{subsec:AHF}
The Amiga Halo Finder \citep[AHF;][]{AHF} was applied to Ms.Marvel DMO, Storm CDM DMO and Storm CDM+baryons to identify DM halos, subhalos, and the baryonic content within. AHF utilizes a spherical top-hat collapse technique to calculate each halo’s virial radius and mass. These simulations identify the virial radius as the radius at which the halo density is 200 times the critical density of the universe at a given redshift. Each halo's virial mass ($M_{200}$) is calculated as the mass contained within the virial radius. We limit our sample of Ms.Marvel DMO and Storm CDM DMO halos to halos with a minimum of $10^4$ DM particles, enabling us to spatially resolve the inner structure of CDM down to 0.33 kpc. For a detailed discussion on resolution see Appendix \ref{sec:app_B}
Resolved galaxies from Storm CDM+baryons contain a minimum of 14 star particles and have resolved star formation histories (SFHs) which extend past 100 Myr. We consider a sample of 228 halos from the Ms.Marvel DMO simulation, 235 halos from the Storm CDM DMO simulation, and 20 galaxies from the Storm CDM+baryons simulation. We note that the Storm CDM+baryons contains on the order of 200 resolved DM halos, however, many are completely dark and only 20 halos host galaxies with enough star particles to be considered resolved. The selected halos do not include satellites. 

\subsection{Ms.Marvel DMO}
\label{sec:Ms.Marvel}
\indent Ms.Marvel DMO simulates a $25^3$ Mpc$^3$ volume and uses a DM particle mass of 8070 $M_\odot$ with a gravitational force resolution of 65 pc \citep{Bellovary2019}. Where 65 pc is half the spline kernel width. This high resolution allows DM halos that range in halo mass from $M_{200}\approx8\times10^7M_\odot$ and contain a minimum of $10^4$  DM particles such that their halo centers are well-resolved prior to core-collapse. We show in Appendix \ref{sec:app_B} that this particle resolution is sufficient to resolve CDM halos with a spatial resolution of 0.33 kpc in accordance with \cite{ShrinkSphere}. However, we note that \cite{Mace2024} find a minimum requirement of $10^5$ DM particles to fully resolve core-collapse, indicating that our SIDM halos in the late stages of gravothermal evolution may be subject to resolution effects. 
\newline\indent Here we describe the SIDM model and implementation for the DMO run of Ms.Marvel. Our SIDM model is run with a velocity-dependent cross section with a transfer cross-section for a Yukawa potential \citep{Feng2010,loeb2011} in the classical regime. We present a sample of high-resolution simulated dwarf halos, in a full cosmological volume centered on isolated dwarf galaxies. The velocity-dependence of our self-interaction cross section is implemented to allow a subset of halos to undergo core-collapse within a Hubble time. The velocity-dependence of the cross section is parameterized by a maximum interacting cross section divided by the mass of the dark matter particle equal to $\sigma_{\rm{max}}/m_\chi = 50 $~cm$^2$/g at a given maximum circular velocity $v_{\rm{max}} = 35 $~km/s. The cross section at a given maximum circular velocity is tuned such that the effective cross section of interaction is on the order of $10^1$ cm$^2$/g for classical dwarfs and $10^2$ cm$^2$/g for our lowest-mass ultra-faint dwarfs (UFDs). The implementation of SIDM particle physics follows from previous work by \cite{Tulin2013} and \cite{Tulin18}. See Cruz et al.(in prep) for additional details on the SIDM model and it's implementation. 
\begin{figure}
 \centering
 \includegraphics[width=1\linewidth]{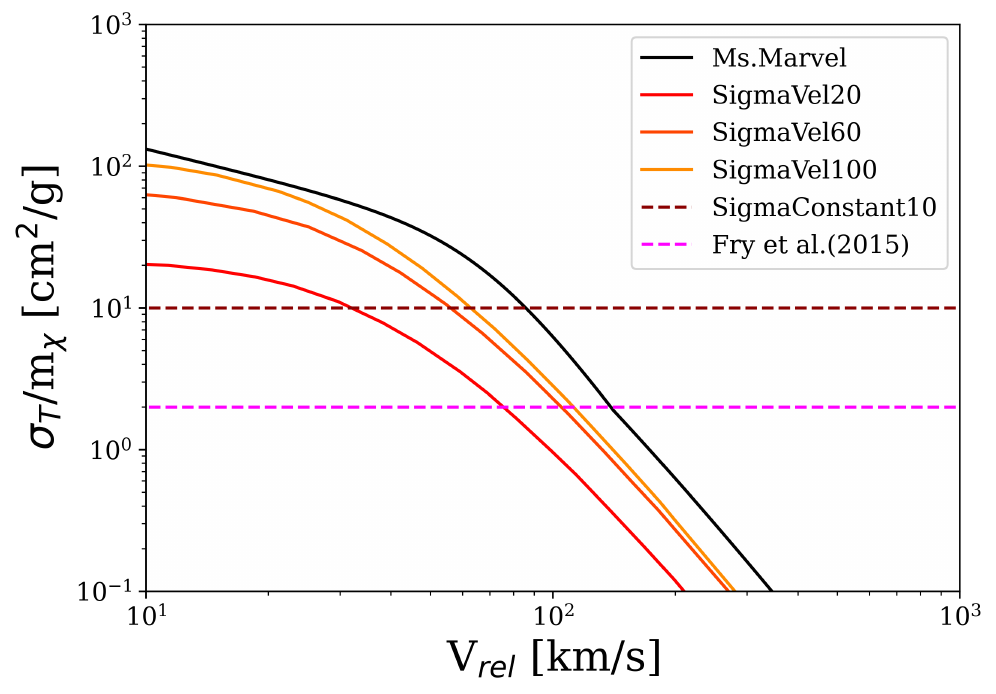}
 \caption{The effective cross section versus $V_\text{max}$ for Ms.Marvel (black), with comparisons from \cite{Fry2015} in pink and \cite{Correa_2022}; SigmaConstant10 (dark red), SigmaVel20 (red), SigmaVel60 (orange red), and SigmaVel100 (orange). Velocity-dependent models are shown with solid lines and models with constant cross sections are shown with dashed lines. We include different models in this plot to contextualize our comparison between the inner slopes of the DM density profiles of dwarf halos from the Ms.Marvel simulation against halos from \cite{Correa_2022} and \cite{Fry2015} in a later section.}
 \label{fig:sigma_eff}
\end{figure}

\indent The transfer cross section of SIDM particles in our simulations is given by the following piecewise function from \cite{Tulin2013}:
\begin{equation}
 \sigma_T = 
 \left \{
 \begin{array}{ccc}
 \frac{4 \pi}{ m_\phi^2} \beta \space \rm{ln}(1 + \beta^{-1}) & \beta \le 10^{-1} & \\ \frac{8 \pi}{ m_\phi^2}\beta^2 (1 + 1.5\space \beta^{1.65})^{-1} & 10^{-1} \le \beta \le 10^3
 & \\ \frac{\pi}{ m_\phi^2} (\rm{ln}\beta + 1 - 0.5\space\rm{ln}^{-1}\beta)^2 & \beta \ge 10^3
 \end{array}
 \right \} 
\end{equation}
where $\beta \equiv 2\alpha_\chi m_\phi/v_{rel}^2m_\chi$, $m_\chi$ is the DM mass, $m_\phi$ is the mass of the light-mediator, $v$ is the relative velocity between DM particles, and the coupling $\alpha_\chi$ is the DM analog of the fine structure constant. We show the transfer cross section for our SIDM model implemented in Ms.Marvel compared against various velocity-dependent models trained on MW-satellites from \cite{Correa_2022} and SIDM with a constant cross section simulated in both a MW environment \citep{Correa_2022} and isolation \cite{Fry2015} in Figure \ref{fig:sigma_eff}. We show velocity-dependent models with solid lines and models with constant cross sections with dashed lines. We compare the inner slope of the DM density profiles of dwarf halos from the Ms.Marvel simulation against the SIDM models from \cite{Correa_2022} and \cite{Fry2015} in Figure \ref{fig:sigma_comp}. 
\newline\indent The SIDM model that we implement in this simulation is a light mediator model with elastic scattering. Simulation particle scattering is stochastic with the probability of particle collision between two DM particles given as a function of relative velocity and time interval at local density with cross section per unit mass of DM interaction:
\begin{equation}
 P = \rho \sigma_{dm} v_{rel} \Delta t \label{eq:delta_t}
\end{equation} 
The timestep is adjusted such that the collision probability meets the criterion of $\leq  .1$ per timestep, such that for characteristic conditions (the stages of gravothermal evolution prior to the core-collapse phase) $\rho\sigma_{dm}\sigma_{vel}\Delta t \leq 0.1$, where $\sigma_{vel}$ is the local velocity dispersion and $\rho$ is the local density of each particle. The velocity dispersion is indicative of the temperature of DM particles. 
\newline\indent The scattering probability is calculated for each particle with its $k = 32$ nearest neighbors with a spline kernel at every time-step. To avoid multiple simultaneous scattering events the probability of a particle collision between two DM particles is set to be much smaller than once per time step. We then solve Equation \ref{eq:delta_t} to set $\Delta t$ for the simulation. When DM particle collisions occur, the scattering is simulated by a random reorientation in velocity in the center-of-mass frame of the collision, with conserved energy and momentum. The particle's new momentum direction is then randomly assigned. As a result, SIDM halos form isotropic central cores.
\newline\indent Modeling DM halos in the advanced stages of core-collapse presents many unique challenges to N-body simulations \citep{YangYu2022,Zhong2023,Mace2024,Palubski2024,Fischer_2024,fischer2025}. For instance, \cite{zeng2025tillcorecollapsesevolution} find that the central density of core-collapsed halos increases by at least a factor of 5, necessitating a prohibitively small time-step size. In the case where time-steps are too large it is possible for particles to undergo multiple scatterings in a single time-step, leading to errors in energy conservation \citep{Robertson2017,Fischer_2024,zhang2024gd1stellarstreamperturber}. Though \cite{Fischer_2023} show that these errors in energy conservation can be mitigated by the implementation of an algorithm that explicitly handles multiple scatterings. These issues are most prominent in collapsed halos where extremely high central densities induced by the gravothermal catastrophe increase the rate of scattering events. However, \cite{fischer2025} find that using time-steps which are time symmetric can reduce errors in energy conservation. In this work we do not perform special treatment in core-collapsing halos due to the low number of dwarfs that reach core-collapse. However, we note that, as a result, the central densities reached by core-collapsing halos should be treated as a lower limit. 
\newline\indent Other simulation parameters such as the gravitational softening length and mass resolution of the DM particle can affect core-collapse timescales. Accurately simulating core-collapse requires high particle resolution as shown by \cite{Mace2024} which find that collapse times are subject to resolution effects in halos with fewer than $10^5$ DM particles. They find that affected halos have large variations in their collapse times that are slightly biased towards longer core-collapse timescales. Similarly, \cite{Palubski2024} find variations in the collapse times of simulated SIDM halos on the order of 10\% in halos containing as many as $10^6$ particles. \cite{Palubski2024} also find that adaptive gravitational softening artificially accelerates the gravothermal evolution by cooling the SIDM halo. \cite{fischer2025} suggest that these errors in energy conservation arise from gravitational softening lengths which are too small. They find that, if the softening length lies within a reasonable range, halo evolution is insensitive to this simulation parameter. The gravitational softening length used in this work lies above the minimum recommended threshold in \cite{ShrinkSphere}, indicating that halo evolution should be largely insensitive to this parameter in the Ms.Marvel simulations. 

\subsection{Storm CDM DMO}
\indent To compare to predictions from CDM, we include a sample from a DMO run of Storm. These simulations have DM particle masses of 8070 $M_\odot$, and a force resolution of 65 pc. As before, this high resolution allows halos in the range of $M_{200}\approx8\times10^7M_\odot$ to $10^{11}M_\odot$ to be considered resolved down to 0.33 kpc. 

\subsection{Storm CDM+baryons}
To explore feedback-driven core-formation we include a sample of galaxies from a baryonic run of the Storm CDM simulation from the Marvel suite. Storm CDM+baryons is run with the same initial DM conditions as Storm DMO, but includes baryonic particles and additional subgrid physics models for baryonic physics processes such as star formation and stellar feedback. These simulations implement gas, initial star, and DM particle masses of 1410 $M_\odot$, 420 $M_\odot$, and 6650 $M_\odot$, respectively, and have a force resolution of 60 pc. We consider galaxies with $M_*\gtrsim 3\times10^3M_\odot$ to be resolved. 
\newline\indent Storm CDM+baryons implements superbubble feedback \citep{Keller2014} treatment for type II SNe, type Ia SNe, and stellar wind. At the time of the supernova, mass, metals and energy are directly injected into surrounding gas particles. In the subsequent time-steps, subgrid processes for thermal conduction and evaporation regulate the quantity of SN heated gas. To treat superbubbles with initial sizes that lie below the gas particle mass resolution of our simulations, we adopt a two-phase treatment of gas particles, which includes a hot phase and a cold phase. This two-phase treatment allows thermal conduction to take place between the hot and cold phases of SNe heated gas particles until the hot phase is no longer very hot ($ \sim 10^5 K$)
\newline\indent We implement a stochastic star formation recipe with a star formation efficiency which is dependent on the local non-equilibrium abundance of molecular hydrogen along with gas cooling from \cite{Christensen12}. Star formation takes place in particles that are sufficiently cold ($T<10^3K$) and dense ($n>0.1 $ amu cm$^{-3}$) though stars typically form above $100 $ amu cm$^{-3}$ because of the H$_2$ requirement. The probability of a star particle forming within a time $\Delta t$ is given by:
\begin{equation}
 p = \frac{m_\text{gas}}{m_\text{star}}(1-e^{-c_0^*X_{H_2}\Delta t/t_\text{form}})
\end{equation}
where $X_{H_2}$ is the mass fraction of Hydrogen in the form of $H_2$. In the equation above $m_\text{gas}$ is the mass of the gas particle, $m_\text{star}$ is the initial mass of the star particle that forms, and $t_\text{form}$ is the local dynamical time. The star formation efficiency parameter ($c_0^*$) is a tunable parameter set to 0.1 such that star formation follows the normalization of the Kennicutt-Schmidt relation \citep{Christensen14}. Star particles represent a population of stars born with a \cite{Kroupa01} initial mass function. 

\section{Results}
\label{sec:results}
\indent We organize this section as follows. In Section \ref{sec:z=0} we measure the inner density slopes of halos from all three of our simulations (Ms.Marvel SIDM DMO, Storm CDM DMO, Storm CDM+baryons) at redshift zero, then trace both the central densities and inner density slopes of these halos through time to study core formation. We isolate a sample of core-collapsed SIDM halos from Ms.Marvel DMO in Section \ref{subsec:trace}, and identify a threshold for core-collapse in Section \ref{subsec:collapsed}. We compare core-collapse timescales with predictions from a parametric model of gravothermal evolution \citep{Yang2023} in Section \ref{subsec:param}.

\subsection{Core formation in SIDM and CDM simulations}
\label{sec:z=0}
\begin{figure*}
 \centering
 \includegraphics[width=1\linewidth]{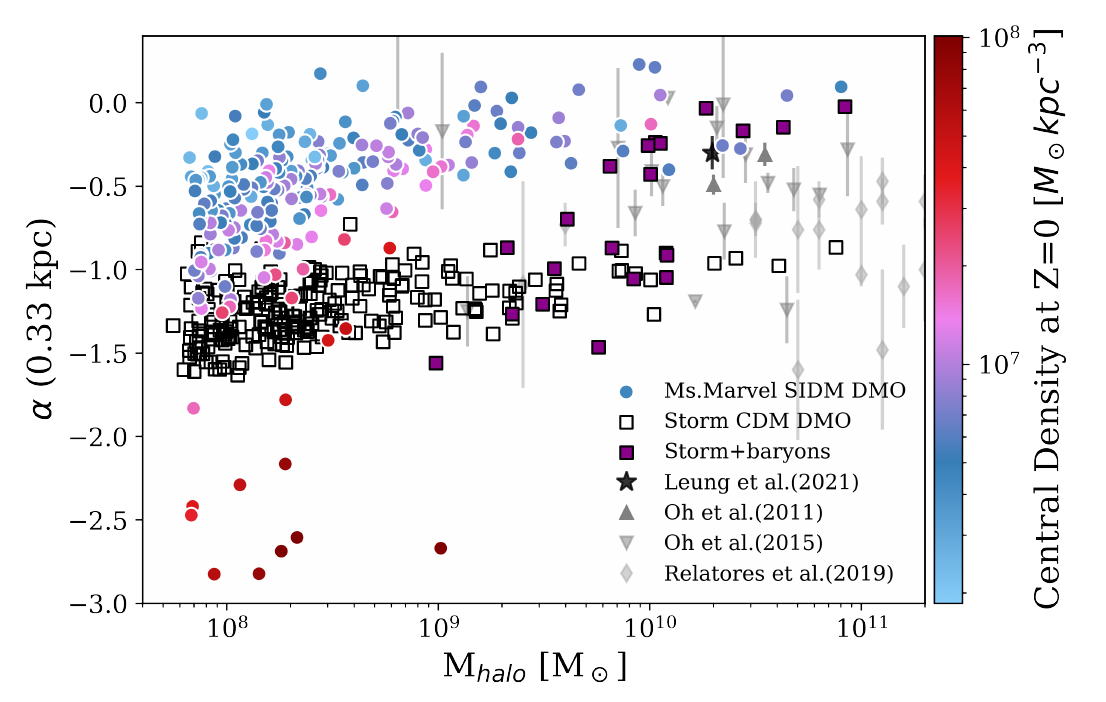}
 \caption{Central inner density slope ($\alpha$) vs DM halo mass colored by central density. Colored circles, empty squares, and purple squares represent halos from the Ms.Marvel DMO, Storm CDM DMO, and Storm CDM+baryons simulations respectively. Grey points represent observed slopes in dwarf galaxies from \cite{Leung2021} (square), \cite{Oh2011slopes} (up triangle), \cite{Oh2015} (down triangle), and \cite{Relatores2019} (diamond). The inner density slope of a smoothed DM density profile is measured between 0.26 and 0.39 kpc in each halo. Ms.Marvel halos are colored by corresponding central density, calculated as the averaged density within a radius of 0.33 kpc.}
 \label{fig:Mvalpha}
\end{figure*}

\indent As discussed in the introduction, previous works have found that core-formation in $\Lambda$CDM can be achieved through the inclusion of strong baryonic feedback, while SIDM halos naturally produce central density cores. We directly compare core formation between SIDM and CDM simulations with our Ms.Marvel DMO and Storm CDM+baryons halos in Figure \ref{fig:Mvalpha}, which plots inner density slope against halo mass. We include halos from our Storm CDM DMO simulations, whose unaltered DM density profiles serve as a baseline comparison. Ms.Marvel DMO halos are represented by circles colored by their $z=0$ central density, while Storm CDM DMO and Storm CDM+baryons halos are marked by empty squares and purple squares respectively. Coloring SIDM halos by their central density illustrates the relationship between central density and the slope of the DM density profile at the center of each halo. Central density is calculated as the average density contained within a 0.33 kpc radius from the center of the halo. We find that above halo masses of $\sim10^{10}M_\odot$, both DM self-interactions in SIDM halos and baryonic feedback in CDM halos produce central density cores. At lower halo masses, only SIDM halos can form cores.
\newline\indent We measure inner density slope at 0.33 kpc ($\sim 5 \times$ the softening length) by applying a linear fit in log-log space to the DM density profile of the halo calculated through {\it{pynbody}} \citep{Pontzen13} between 0.26 and 0.39 kpc. \footnote{Measuring at a fixed radius means that the inner density slope is measured at larger distances (with respect to the virial radius) as mass decreases. As a result, in Figure \ref{fig:Mvalpha} the slopes of the density profiles steepen with decreasing halo mass.} We chose this radius because it is the smallest radius where the DM density profile of the halo remains spatially resolved. We note that we measure the slope of the DM density profile outside of the core radius for all but the most massive halos in our sample, however, as we show in figure \ref{fig:Mvalpha} we are still able to differentiate between core-forming SIDM halos and core-collapsing SIDM halos at this radius.
The DM density profiles are calculated from the mass profiles generated by Tangos, a system that creates and organizes databases of galaxy and halo properties in simulations \citep{Pontzen_2018_tangos}. We find that, for SIDM halos in particular, this practice best preserves the high central densities of DM at the innermost radii. Additionally, compared to a differential measurement of the density profile, using enclosed mass at each radial bin to calculate the average density yields less noise.

\indent The differences in the inner density slopes of low-mass ($M_{halo} <10^{10}M_\odot$) SIDM and CDM halos results from how cores are formed differently between these two models of DM. Previous work by \cite{Bursting2024} studying the Storm CDM+baryons simulations, shows that in $\Lambda$CDM cores are formed and sustained through repeated feedback events. As a result, in low-mass halos, which have fewer stars and less extended star formation histories (SFHs), baryonic feedback is insufficient to convert cusps into cores. Both \cite{Bursting2024} and \cite{Lazar2020} find that the slopes of UFDs, which are defined by \cite{Lazar2020} as galaxies with $M_*/M_{halo} \leq 5 \times 10^{-5}M_\odot$, are indistinguishable from DMO halos. In this work, we define UFDs as halos less massive than $10^{9} M_\odot$.\footnote{UFDs are typically defined as galaxies with $M_* < 10^5 M_\odot$ \citep{Bullock17,SantosSantos22}. \cite{Bursting2024} find that, in the Storm simulation, galaxies at the upper end of the UFD mass range are hosted by halos with $M_{200} \simeq 10^9 M_\odot$. These results are in agreement with previous published stellar-mass-halo-mass relations \citep{Jethwa18,Nadler20}.} Figure \ref{fig:Mvalpha} shows that as halo mass decreases, the inner density slopes of Storm CDM+baryons halos become steeper, converging to the NFW profile predicted by $\Lambda$CDM DMO simulations. In the Storm CDM DMO run, halos remain cuspy at all halo masses, with inner density slopes ranging between $\alpha \sim -1.7$ and $-1$, despite a wide range of central densities. In contrast, Ms.Marvel DMO halos form cores through DM self-interactions. We find that these Ms.Marvel DMO halos with $M_{halo} < 10^{10} M_\odot$ produce DM density profiles that are significantly more cored relative to Storm CDM+baryons halos of the same mass. Because SIDM halos with sufficiently small initial concentrations can have extremely long core-collapse timescales, even with large interacting cross sections, we expect that velocity-dependent SIDM models that naturally produce cores in the bright dwarf range will also produce some cored halos at lower masses. 
\newline\indent Although we find evidence for core formation across the entire mass range of Ms.Marvel DMO halos, 9 of the Ms. Marvel DMO halos have steeper inner density slopes, $\alpha < -2.0$, exceeding the steepest measured slope from our Storm CDM DMO counterparts. These steeply cusped SIDM halos have increased central densities that in some cases exceed the central densities of cusped Storm CDM DMO halos, which are over 10 times more massive. We find that Ms.Marvel halos are only this steeply cusped at halo masses below $M_{halo}\approx 10^9 M_\odot$. In Section \ref{subsec:param} we show that halos more massive than $M_{halo} = 10^9 M_\odot$ have core-collapse timescales that exceed a Hubble time and remain cored. These results indicate that both the median inner density slope and the scatter in inner density slope as a function of halo mass of Ms.Marvel DMO halos exhibit a strong mass dependence, which is expected given the implementation of a velocity-dependent cross section in our SIDM model. 
\newline\indent In Figure \ref{fig:Mvalpha}, we compare with observed central DM density slopes (shown in grey) from \cite{Oh2011slopes}, \cite{Oh2015} , \cite{Relatores2019} , and \cite{Leung2021}. The observed DM density profiles are calculated from rotation curves assembled using gas tracers \citep{Relatores2019}, stellar kinematics \citep{Leung2021}, or a mix of baryonic components \citep{Oh2011slopes,Oh2015}. \cite{Relatores2019} measure the slope of the density profile between 300-800 pc, while \cite{Oh2011slopes} and \cite{Oh2015} measure slopes between 0.1-0.2 kpc. \cite{Oh2011slopes} and \cite{Oh2015} calculate the dark matter density profile from measurements of the rotation curve using the Poisson equation under the assumption of a spherical mass distribution. They trace the motion of the gas component and use the Gauses-Hermite polynomial to model the velocity field. The error bar in the rotation velocities indicates the dispersion of individual velocity values found along a tilted ring. \cite{Relatores2019} extracts rotation curves using DiskFit to model the fields produced from the Palomar Cosmic Web Imager data, the error to the fit is represented by errorbars. For dwarf galaxies with resolved stellar kinematics and well behaved HI gas rotation curves, \cite{Leung2021} jointly model the stellar and gaseous kinematics. They use the HI integrated intensity map and a discrete set of line-of-sight velocity measurements from 180 member giant branch stars to calculate the circular velocity of WLM.
\cite{Oh2011slopes} and \cite{Leung2021} study a set of isolated field dwarfs with little to no gravitational interactions with other galaxies. Both data sets from \cite{Relatores2019} and \cite{Oh2015} include dwarf galaxies in a range of environments, however we exclude known satellites for an apples-to-apples comparison between observations and the isolated field dwarfs from the Ms.Marvel DMO and Storm CDM DMO simulations. SIDM satellites are also subject to host halo effects such as tidal and ram-pressure stripping, which can impact their evolution~\citep{Nadler:2020ulu}. For example, tidal interactions have been shown to disrupt the gravothermal evolution of simulated SIDM satellites \citep{Carton2022,Jiang19,Benavides2021,Benavides2023}. However, \cite{Nishikawa2020} and \cite{Carton2022} find that tidally stripped SIDM halos can also have accelerated core-collapse timescales. In the $\Lambda$CDM paradigm tidal stripping has been shown to lower the central DM densities, particularly in cored satellites \citep{Zolotov12,brooks2014}. 
\newline\indent The substantial differences in the predicted slopes of the DM density profiles at halo centers of classical dwarf galaxy mass halos ($10^{9} M_\odot -10^{10} M_\odot$) could be used to discriminate between DM models observationally. Of the 4 lowest-mass observed dwarfs (IC 1613 \citep{Oh2015}, CVnIdwA \citep{Oh2015}, DDO 216 \citep{Oh2015}, and NGC 949 \citep{Relatores2019}), two are distinctly cuspy while the others are more cored, with one positive core slope measurement. Strong, repeated bursts of stellar feedback are required to form cores in CDM halos \citep{Bursting2024,Mostow_2024_bursts}, as a result, simulations predict that CDM halos in this mass range should have cuspy inner DM density profiles. In other words, cored low-mass halos are incompatible with current predictions from $\Lambda$CDM simulations. In contrast, we find that the slopes of Ms.Marvel DMO halos can fall within a range of values, from cored to steeply cusped. However, observed isolated dwarfs from \cite{Oh2011slopes}, \cite{Oh2015} , \cite{Relatores2019} , and \cite{Leung2021} do not prefer either DM model shown in this work. 
\newline\indent By measuring the inner density slopes of low-mass dwarfs discovered in new and upcoming surveys or finding ways to decrease the uncertainty of this measurement, we have the potential to rule out or place constraints on some DM models. 

\begin{figure*}
\centering
 \includegraphics[width=1\linewidth]{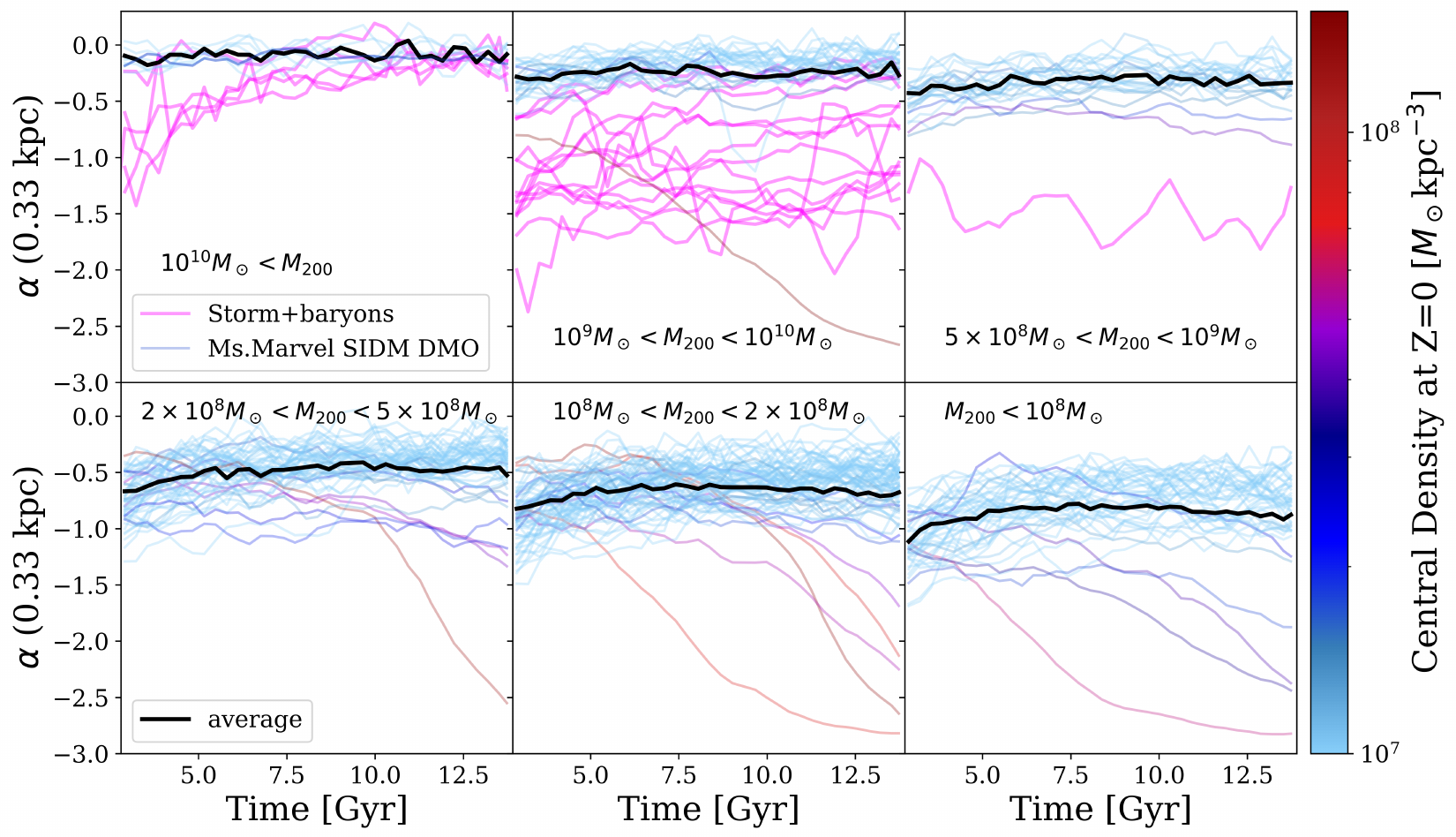}
 \caption{Central inner density slope traced through time for Ms.Marvel DMO halos. Inner density slope of a smoothed DM density profile is measured between 0.26 and 0.39 kpc in each halo. Each solid line represents one halo, and lines are colored by halo central density at redshift zero. Halos are separated into different panels by halo mass measured at redshift zero, decreasing from left to right and from top to bottom. The black line represents the average slope traced through time for the halos in each mass bin. Magenta lines represent slopes traced through time for Storm CDM+baryons halos. Inner density slopes are measured using the same methodology as in Figure \ref{fig:Mvalpha}. }
 \label{fig:sidm_trace_alpha}
\end{figure*}
\begin{figure*}
\centering
 \includegraphics[width=1\linewidth]{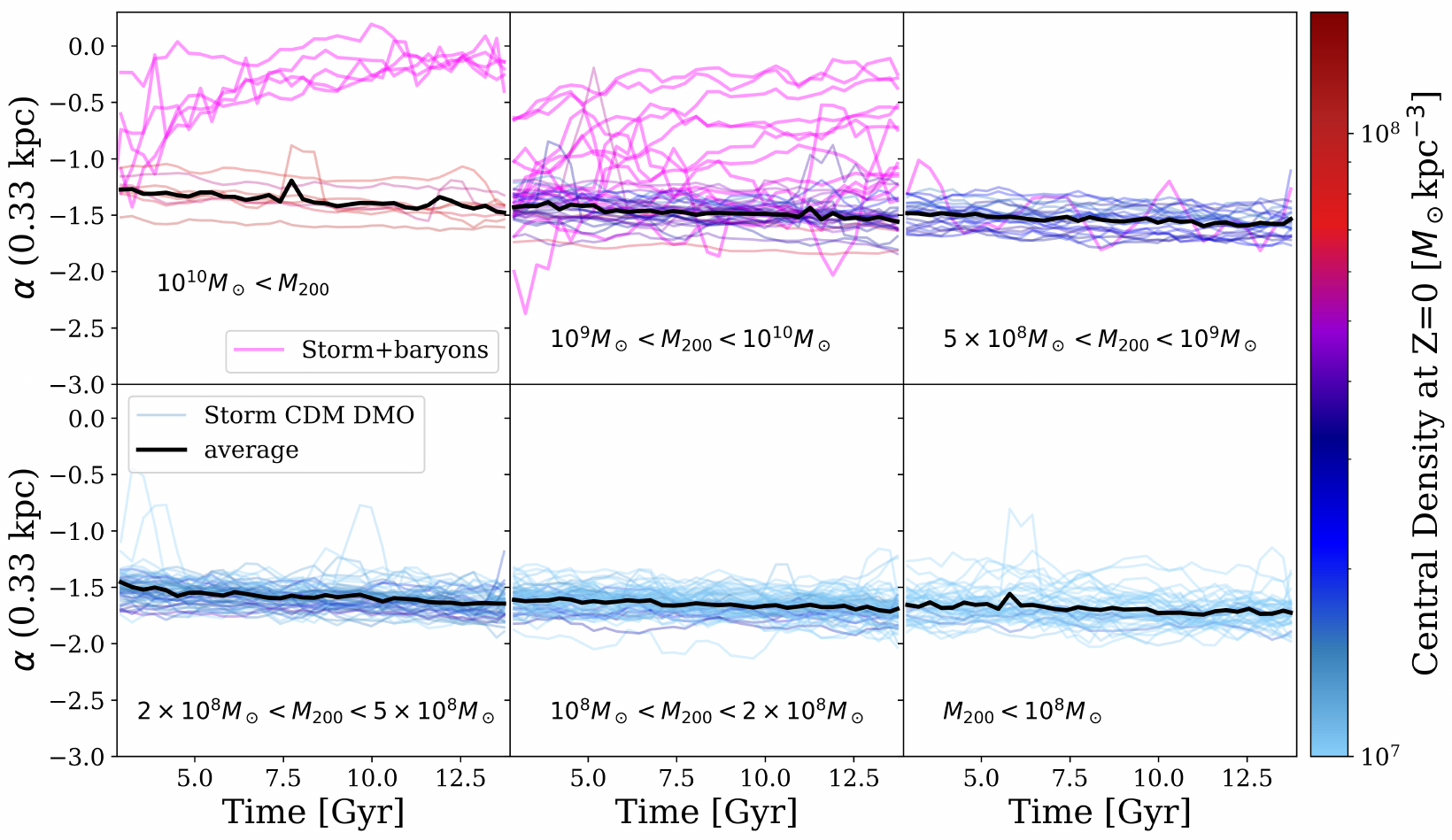}
 \caption{Central inner density slope traced through time for Storm CDM DMO halos. inner density slope of a smoothed DM density profile is measured between 0.26 and 0.39 kpc in each halo. Each solid line represents one halo, and lines are colored by central density at redshift zero. Halos are separated into different panels by halo mass measured at redshift zero, decreasing from left to right and from top to bottom. The black line represents the average slope traced through time for the halos in each mas bin. Magenta lines represent slopes traced through time for Storm CDM+baryons halos. Inner density slopes are measured using the same methodology as in Figure \ref{fig:Mvalpha}. }
 \label{fig:cdm_trace_alpha}
\end{figure*}

\subsection{Tracing inner density slope and central density through time}
\label{subsec:trace}
\indent Tracing inner density slope and central density over time enables us to study the formation of cores in Ms.Marvel DMO halos and compare their evolution with their Storm CDM+baryons counterparts. Through this process, we can visually identify core-formation and potential core-collapse in the Ms.Marvel DMO run. We trace the inner density slopes of each Ms.Marvel DMO and Storm CDM+baryons halo in Figure \ref{fig:sidm_trace_alpha} while Figure \ref{fig:sidm_trace_den} traces the growth in central density between 2.26 and 13.78 Gyr, normalized by each halo's initial central density for each matched run. In Figures \ref{fig:cdm_trace_alpha} and \ref{fig:cdm_trace_den} we trace properties from the Storm CDM DMO simulation to serve as a benchmark comparison. In the following section, we show that SIDM halos form central density cores at early times, CDM halos with baryonic feedback form cores in higher mass dwarfs at relatively later times, and CDM DMO halos maintain cuspy, NFW profiles throughout their lifetimes.
\newline\indent Figures \ref{fig:sidm_trace_alpha}, \ref{fig:cdm_trace_alpha}, \ref{fig:sidm_trace_den}, and \ref{fig:cdm_trace_den} are arranged as follows; each figure contains six separate panels that are divided into different mass bins, ordered from high to low mass, going from left to right and top to bottom. Each solid line represents the smoothed profile of an individual halo and is colored by its central density at redshift zero. The bluer lines correspond to decreased densities at redshift zero. The central density at redshift zero is calculated using the same methodology described in Section \ref{sec:z=0}. The black line represents the average evolution track for halos in the corresponding mass bin. Plots are smoothed using a low pass infinite impulse response (IIR) filter, described by a differential equation which relies on coefficients to reject high frequencies and pass low frequencies, reducing signal noise. We design a low pass filter using the irfilter function from scipy.signal and apply it to the data set with the scipy.signal.filtfilt function.\footnote{See \hyperlink{scipy.signal}{https://github.com/scipy/scipy/blob/v1.16.1/scipy/signal} for more details}
 \newline\indent While tracing halo properties through time, we find several examples of halos with inner density slopes or central densities that temporarily fluctuate, typically for a few Gyr or less. For example, halo 16 from the Ms.Marvel DMO simulation has a inner density slope of around -0.5 for most cosmic time, but between 7.5 and 12 Gyr the slope dips to nearly $-2.0$. After examining the merger trees from each halo with large fluctuations in either inner density slope or central density, we find that each of these instances coincided with a merger event. Halos with disrupted inner density slopes also show large deviations in central density at the same time-step. We conclude that mergers can temporarily alter the DM density profiles of DM halos \citep{BoylandKolchin2004,Wang2020,Orkney23} for all three simulations; Ms.Marvel SIDM DMO, Storm CDM DMO, and Storm CDM+baryons. The effects of these mergers typically do not exceed 2 Gyr and can significantly disrupt the centers of DM halos during this time. 

\subsubsection{Inner density slopes}
\indent Separating our sample into different mass bins reveals that the evolution of inner density slope is mass-dependent in Ms.Marvel DMO as shown in Figure \ref{fig:sidm_trace_alpha}. We find that all Ms.Marvel DMO halos, excluding those in the lowest mass bin ($M_{\rm halo} < 10^8 M_\odot$), show signs of core-formation with central density slopes shallower than a cuspy NFW profile. However, only the most massive Ms.Marvel DMO halos form true cores, while halos less massive than $10^9 M_\odot$ have cuspier central profiles on average. We define resolved cores and central density cores with core radii that are greater than the spatial resolution used in this work. We measure the core radius of a halo by fitting the DM density profile with a core-Einasto profile \citep{Lazar2020}. We find that Ms.Marvel halos at higher masses are always cored, as far back as we trace halo properties through time. These results indicate that the cores of larger dwarfs ($M_{halo} > 10^9 M_\odot$) are fully formed within the first 2.26 Gyr of the simulation. Figure \ref{fig:sidm_trace_alpha} indicates that while cores form at early times at all halo masses, the average inner density slope reached by redshift zero in Ms.Marvel DMO halos decreases with mass. 
\newline\indent We note that by measuring at a constant radius of 0.33 kpc we measure the inner density slope of lower mass halos at significantly larger radii relative to their virial radii. For halos with masses less than $10^9 M_\odot$ we are likely measuring the inner density slope well outside of the core radius and as a result, find that inner density slopes become steeper with decreasing halo mass. However, we find evidence that most of these halos are likely still core forming by comparing their inner density slopes with the unaltered NFW profiles of Storm CDM DMO halos in Figure \ref{fig:cdm_trace_alpha}. We show that relative to Storm CDM DMO halos, Ms.Marvel SIDM halos are significantly more cored at these lower masses, excepting the cases where the halo is undergoing gravothermal core-collapse which we discuss further in Section \ref{subsec:collapsed}. 
\newline\indent We show in Figure \ref{fig:cdm_trace_alpha} that at all halo masses Storm CDM DMO halos maintain cuspy central profiles throughout their lifetimes that, on average, have inner density slopes between $-1.0$ and $-1.7$. Slopes at the shallower end of this range match predictions from the NFW profile, however, we also find that a significant population of CDM DMO halos have slopes that are steeper. We find that, on average, as halo mass decreases inner slopes become steeper. At lower halo masses the fixed radius of 0.33 kpc we use to measure slope becomes a larger fraction of the scale radius, where the slope of an NFW halo becomes $-2$. As a result we would expect halos at lower masses to have slopes that are steeper than $-1$. 
\newline\indent We note that even in the highest mass range we identify halos with inner density slopes steeper than $-1.5$. Previous works indicate that the NFW profile may underestimate the density of dark matter at the centers of halos \citep{Moore99,Fukushige_2001,Navarro2004}. However, N-body simulations find a wide range of slopes for the central density cusps of CDM halos, some of which are shallower than predictions from the NFW profile \citep{Subramanian_2000_structure, Taylor_2001_phase,ricotti_2003_dependence}. Major mergers \citep{Ogiya_2016,Angulo_2017} and rapid mass accretion \citep{Delos_2022} have been shown to transition steeper cusps into cores. Additionally, \cite{dalal2010origindarkmatterhalo} and \cite{Ludlow_2013_Mass} find that the overall mass accretion history of the halo is a good predictor of the slope of the DM density profile. 
\newline\indent We trace inner density slopes of Storm CDM+baryons halos through time (magenta lines) in both Figures \ref{fig:sidm_trace_alpha} and \ref{fig:cdm_trace_alpha} to compare core formation timescales in the Ms.Marvel simulations with results from our baryonic CDM Storm simulation. We find that relative to SIDM halos, Storm+baryons halos form cores at later times, only after halos have been subject to sufficient repeated rounds of stellar feedback. In accordance with previous results from \cite{Bursting2024}, core-formation is mass-dependent in CDM halos, with only the most massive dwarfs forming large central density cores. In several of the lowest-mass halos in our sample ($\sim 10^9 M_\odot$) the inner density slopes remain cuspy throughout their lifetimes because the limited stellar feedback is insufficient to form central DM density cores. At these halo masses, we find that the slope of the inner density profile is indistinguishable from our CDM DMO run of Storm. 
\subsubsection{Central densities}
\begin{figure*}
\centering
  \includegraphics[width=1\linewidth]{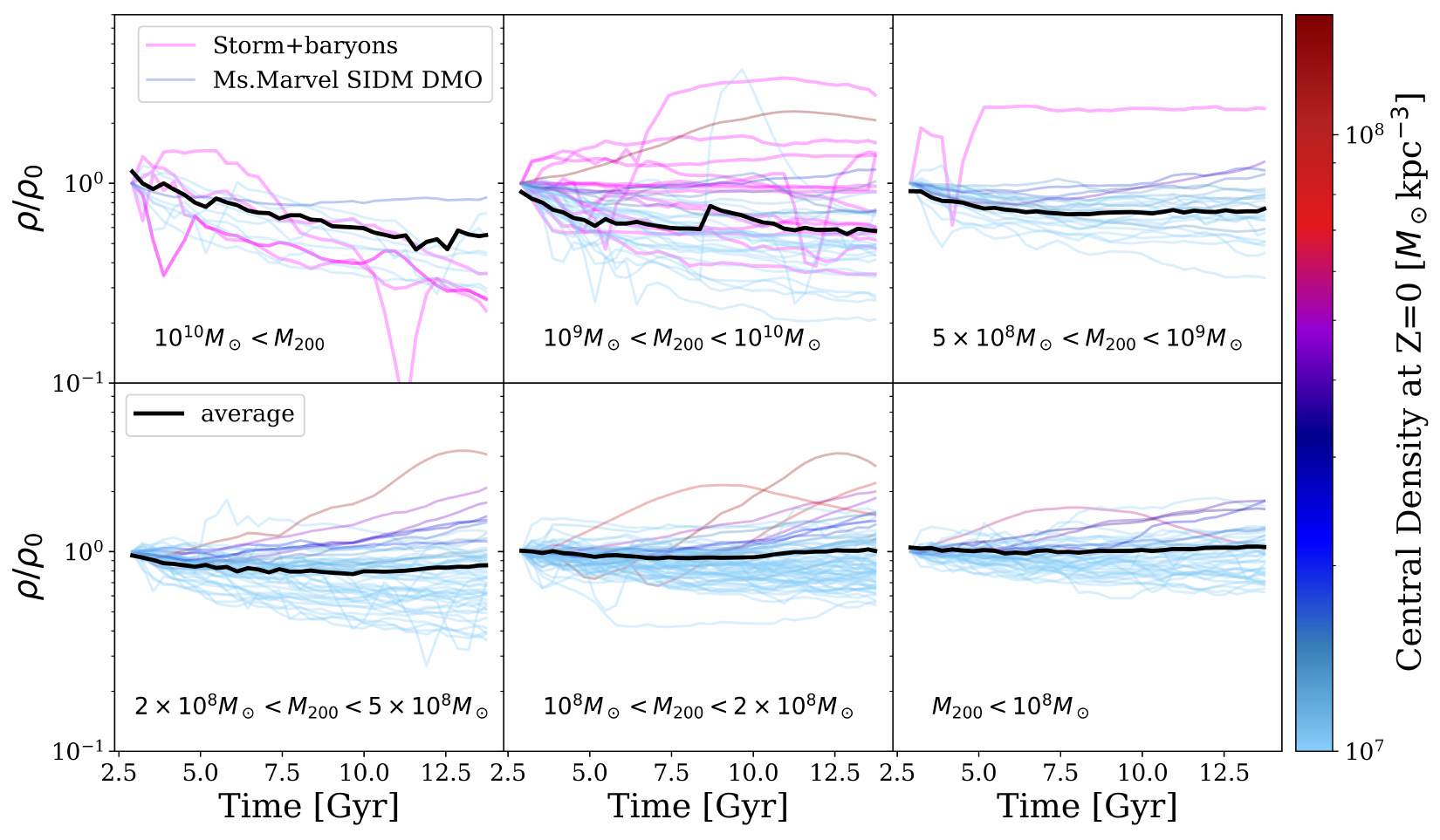}
  \caption{Average central density measured within 0.35 kpc traced through time for Ms.Marvel DMO halos. Central density is normalized by dividing by the initial central density ($\rho_0$) of each halo. Each solid line represents one smoothed halo profile, and lines are colored by central density at redshift zero. Halos are separated into different panels by halo mass with mass decreasing from left to right and from the top to bottom rows. The black line represents the average densities traced through time for the halos in each panel. Magenta lines represent central densities traced through time for Storm CDM+baryons halos. Central densities are calculated by taking the average central density contained within 0.35 kpc of the halo center normalized by initial central density.}
  \label{fig:sidm_trace_den}
\end{figure*}

\begin{figure*}
\centering
  \includegraphics[width=1\linewidth]{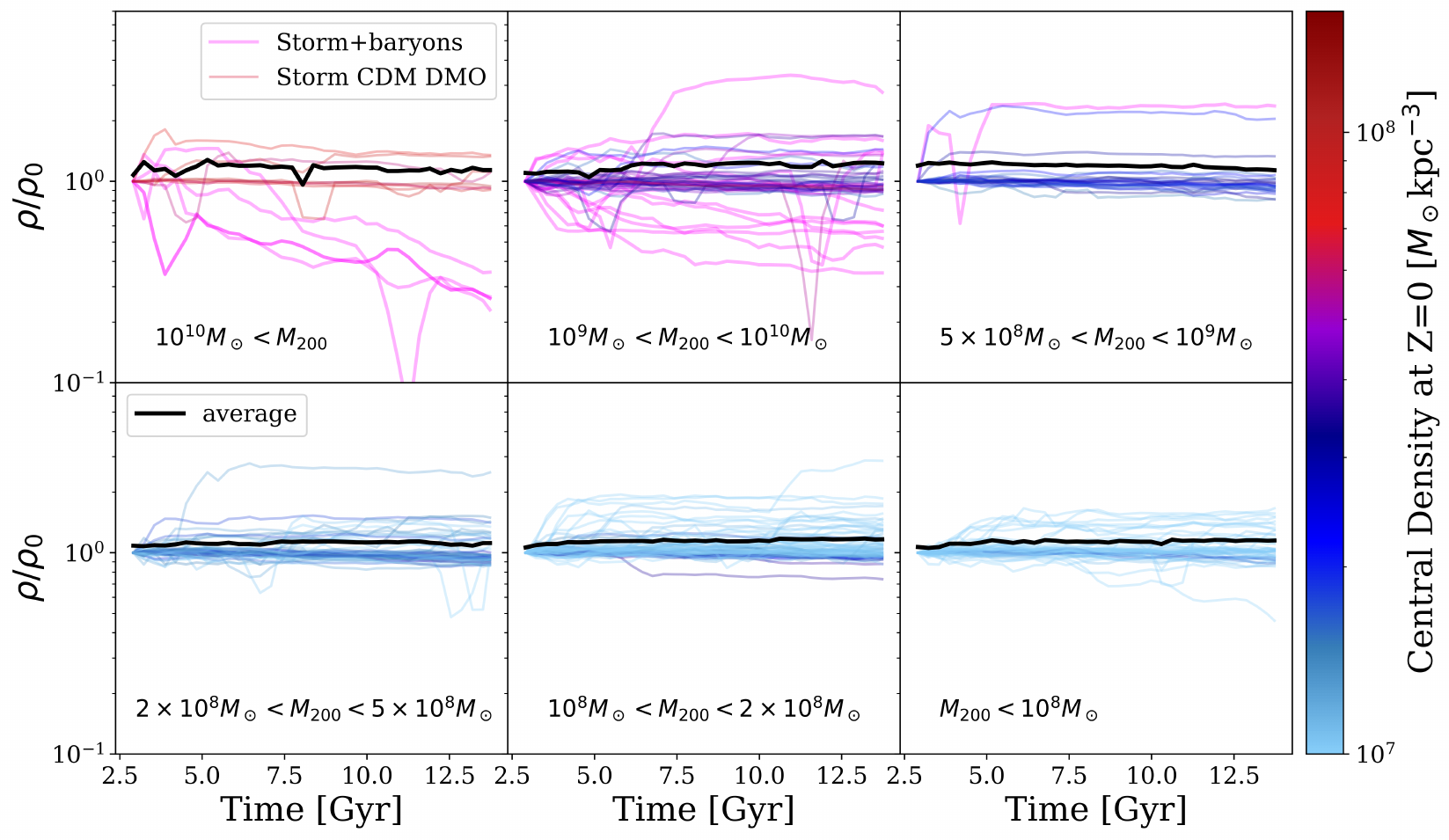}
  \caption{Average central density measured within 0.35 kpc traced through time for Storm CDM DMO halos. Central density is normalized by dividing by the initial central density ($\rho_0$) of each halo. Each solid line represents one smoothed halo profile, and lines are colored by central density at redshift zero. Halos are separated into different panels by halo mass with mass decreasing from left to right and from the top to bottom rows. The black line represents the average densities traced through time for the halos in each panel. Magenta lines represent central densities traced through time for Storm CDM+baryons halos. Central densities are calculated by taking the average central density contained within 0.35 kpc of the halo center normalized by initial central density.}
  \label{fig:cdm_trace_den}
\end{figure*}

\indent By tracing the growth in central density over time for Ms.Marvel halos we show that the evolution of central density tracks inner density slope and, more broadly, core-formation. In Figure \ref{fig:sidm_trace_den}, the highest-mass and most cored Ms.Marvel DMO halos have large central densities at early times that quickly drop to values a fraction of their original value. Most cuspy halos ($\alpha < -1.0$), typically with masses below $5 \times 10^8 M_\odot$, have nearly constant central densities throughout their lifetimes. Results from Figure \ref{fig:sidm_trace_alpha} indicate that cores of the more massive Ms.Marvel halos ($M_{halo} > 10^9 M_\odot$) with the longest core-formation timescales, have well-developed cores by 2.26 Gyr. It is possible that the central densities of halos with lower masses, which typically form cores faster than halos at higher masses, are relatively constant because they have fully formed core at 2.26 Gyr. These results reveal that the evolution of central density as well as inner density slope is mass-dependent in Ms.Marvel DMO. 
\newline\indent We note that the mass-dependence of central density may also be impacted by our choice to measure both the central densities and the inner density slopes of each halo at a fixed radius, regardless of halo mass. When measuring the sizes of cores in these low-mass Ms.Marvel DMO halos, we find that only the most massive halos in our sample have core radii that are larger than the fixed radius where we measure density slopes. Because the core radii of lower-mass halos are smaller than radius of measurement, at these mass scales we measure the slope of the profile outside of the central DM density core. As DM halos form cores, DM particles are pushed out of the inner parts of the halo to larger radii, increasing the density of DM outside of the core. This effect may wash out central density measurements taken far enough outside of the central DM density core. By measuring the central density and the slope of the density profile of lower mass halos ($M_{halo} < 10^9 M_\odot$) at smaller radii we could trace their evolution within the core radius, which would potentially reduce mass-dependence. 
\newline\indent We find that the evolution of central density closely follows inner density slope for CDM halos from Storm CDM DMO and Storm CDM+baryons as well as SIDM halos from Ms.Marvel. The Storm CDM DMO halos shown in Figure \ref{fig:cdm_trace_den} are centrally dense at all times, with very few halos that experience a significant long-term increase or decrease in central density. Similarly to low-mass Ms.Marvel DMO halos ($M_{halo} < 10^9 M_\odot$), the average central density for Storm CDM DMO halos is nearly constant, with a slight increase over time (likely a result of  fast accretion halo growth dominated by major mergers \citep{Zhao_2003,Lu_2006}), independent of halo mass. The baryonic CDM halos from our Storm+baryons simulation colored in dark blue for both Figures \ref{fig:sidm_trace_den} and \ref{fig:cdm_trace_den} decrease in central density over time in the highest mass bin and show a range of behavior at lower masses. The lowest-mass Storm+baryons halo has a central density that remain largely constant over time, mirroring the behavior of Storm DMO halos. We note that the initial growth in central density and subsequent fluctuations in are caused by halo mergers that take place at early times. Between halo masses of $10^{9-10} M_\odot$ the more massive baryonic CDM halos have decreasing central densities over time, closely following the behavior of core-forming SIDM halos from the Ms.Marvel simulation. 


\subsection{Core-collapsed Ms.Marvel DMO halos} \label{subsec:collapsed}
\begin{figure*}
\centering
  \includegraphics[width=1\linewidth]{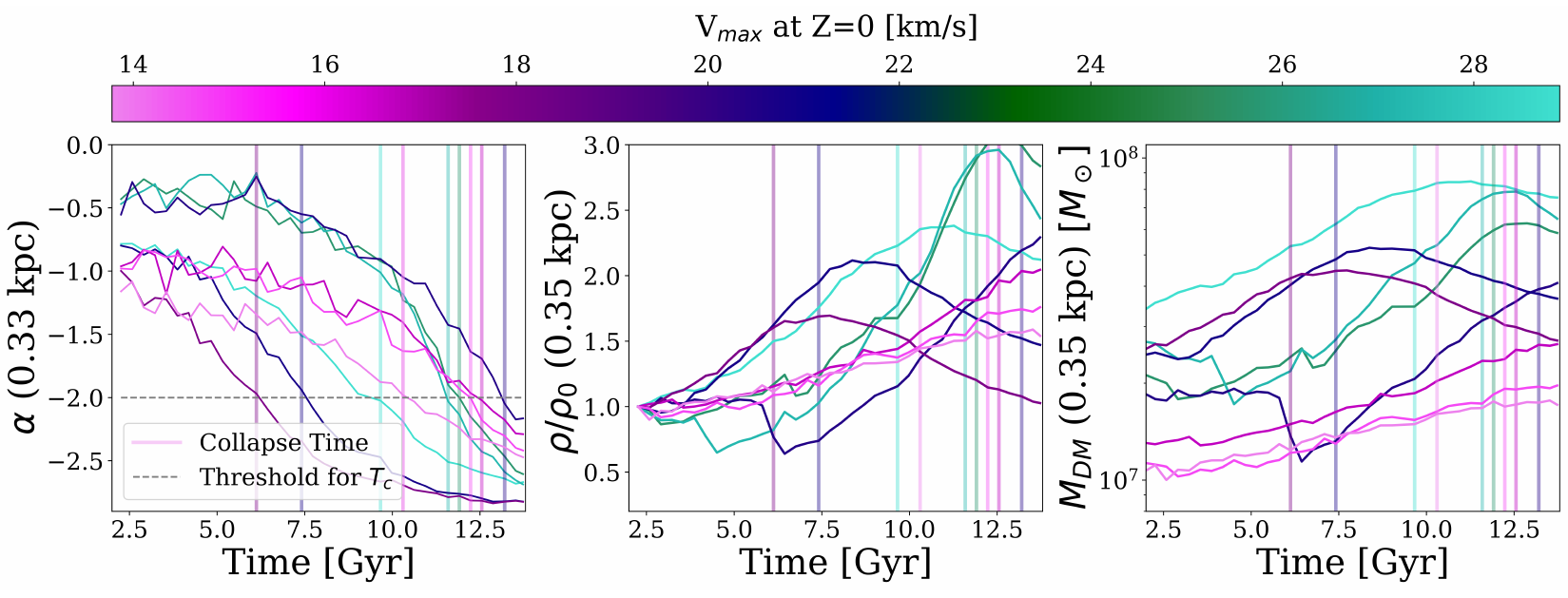}
  \caption{Left panel: Inner density slope measured at 0.33 kpc traced through time [Gyr] for our sample of core-collapsed Ms.Marvel DMO halos. Each solid line represents one halo, and lines are colored by maximum circular velocity with larger $V_\text{max}$ appearing bluer and lower values appearing pinker. Vertical dashed lines intersect with inner slopes at the time when the halo's inner density slope first reaches $-2.0$ and the halo enters the initial stages of core-collapse. The $-2.0$ threshold of inner density slope is shown by the horizontal gray line.
  Middle panel: average central density enclosed within 0.35 kpc for core-collapsed halos traced through time. Central density is normalized by dividing by initial central density. 
  Right panel: Mass enclosed within 0.35 kpc for core-collapsed halos traced through time.}
  \label{fig:trace_core}
\end{figure*}

\indent Core-collapsed halos are characterized by an initial core-forming phase, and later undergo core-collapse. The core-formation and core-collapse timescales are then dependent on the quantity of SIDM self-scattering events. Increasing the size of the cross section leads to more frequent scattering events, which has been shown in previous work to cause SIDM halos to collapse at earlier times \citep{zeng2025tillcorecollapsesevolution}. We analyze the gravothermal evolution of SIDM halos in the Ms.Marvel simulation by tracking the distribution of DM at the center of collapsing halos over time. 
\newline\indent In Figures \ref{fig:sidm_trace_alpha} and \ref{fig:sidm_trace_den}, we identify 13 halos with divergent behavior. This subset of halos is characterized by density profiles that grow cuspier at late times, while the average Ms.Marvel DMO halo grows more cored (at higher masses $M_{halo} > 10^9 M_\odot$) or maintains a constant central density (at lower masses $M_{halo} < 10^9 M_\odot$). In addition to having steeper central slopes, these halos are also more centrally dense at late times. Seven halos have particularly high central densities, exceeding $5 \times 10^7 M_\odot$ kpc$^{-3}$, which is approximately twice the average central density of our Ms.Marvel DMO halos. These highly centrally dense halos have masses $\lesssim 10^9~M_\odot$ and display steadily decreasing inner density slopes for the majority of their lifetimes. In some cases, inner density slopes reach values as low as $-2.5$ which is substantially lower than $-$1.9, the steepest slope identified in our sample of Storm CDM DMO halos. The steepening of the density slopes of Ms.Marvel DMO halos could result from core-collapse. We caution, however, that the resolution limits of the simulations do not permit us to fully resolve core-formation in the mass range of the affected halos. We explore this further in Figure \ref{fig:trace_core}, which isolates a subset of potentially core-collapsed halos. 
\newline\indent In the following section, we define our sample of core-collapsed halos as having inner density slopes that are more negative than $-2.0$ at $z=0$. We track the gravothermal evolution of these Ms.Marvel DMO halos through time in Figure \ref{fig:trace_core} to better illustrate how DM behaves at the centers of halos during core-collapse, and measure core-collapse timescales. The left panel of Figure \ref{fig:trace_core} traces the inner density slopes of 9 core-collapsed halos over time, with each halo colored by its maximum circular velocity. We define our core-collapsed sample as halos with slopes which are more negative than $-2.0$ at redshift zero. We adopt this threshold because it is just outside the range of inner density slope values we identified in our sample of Storm CDM DMO halos. We further define a core-collapse time, $T_c$, where the inner density slope of the halo first drops below a threshold of $-2.0$, illustrated by the vertical dotted lines. We use this collapse time to mark when the SIDM halo enters the initial stages of core-collapse, and is not yet fully core-collapsed. This threshold marks the point when an SIDM halo first becomes more steeply sloped than simulated halos from Storm CDM DMO. In simulations with fully resolved core-collapse, the inner density slope of a fully core-collapsed halo will become significantly steeper than this threshold in both isolated gravothermal simulations \citep{Lynden-Bell1968,Kochanek2000,Balberg2002,Koda2011,Pollack2015,Essig2019,Sameie2020,Correa2021,Turner2021,Carton2022,Tran2024,Mace2024,Fishcer_2024_Cosmological} and cosmological simulations \citep{Elbert2015,Essig2019,Zavala2019,Nishikawa2020,Palubski2024,Fishcer_2024_Cosmological,Tran2025}. We then test how well our measured collapse time compares to predictions from analytic models. 

\indent The middle and right-hand panels in Figure \ref{fig:trace_core} measure the amount of DM contained within the halo center over time using different properties for direct comparison. The middle panel traces a normalized central density using the same methodology as in Figure \ref{fig:sidm_trace_den} and the right panel traces the total mass contained within 0.35 kpc. Comparing the middle and right-hand panels to the evolution of the inner density slope in the left panel clearly shows how the slope reacts to increases in the density/mass of DM located at the halo center. As expected, we find that the inner density slope is inversely related to the central density, where the sharpest decrease in the density slope corresponds to the highest rate of growth in central density. 
\newline\indent We find that most core-collapsed halos  have fully formed cores by 2.5 Gyr, evidenced by their lowered central densities and shallower inner density slopes. Core-collapse occurs at a variety of times for Ms.Marvel DMO halos, with halos entering the initial stages of core-collapse as early as $\sim6$ Gyr and some not entering the core-collapse phase until nearly the end of the simulation. Additionally, we find that both the central density and enclosed mass are good tracers of core-collapse. Specifically, an increase in either of these quantities can indicate core-collapse. 
\newline\indent Three of the halos in the core-collapsed sample have a slight decrease in central density and enclosed mass at late times. In Appendix \ref{sec:app_A} we show that this unexpected turndown in DM density results from measuring central density too far outside the halo center. With higher spatial resolution, it is possible to more confidently measure halo properties within smaller radii. However, we show that the growth of the central density remains mostly stable down to one gravitational softening length. 

\begin{figure*}
\centering
  \includegraphics[width=1\linewidth]{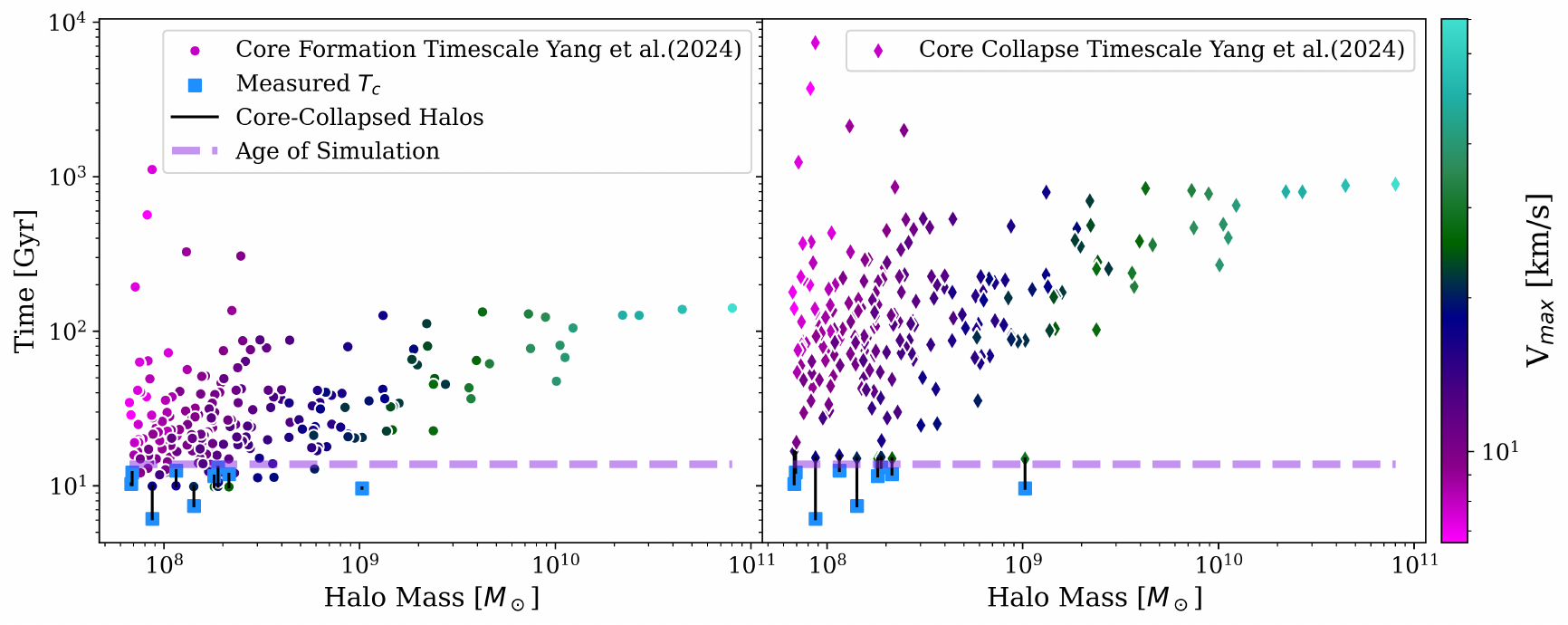}
  \caption{Left side: Predicted core-formation timescales (defined as $\tau = 0.15$) plotted against halo mass for Ms.Marvel DMO halos. Blue squares mark the time when the inner density slope of a Ms.Marvel DMO halo first drops below $-2.0$ while colored circles represent the predicted core-formation timescales, calculated analytically from a parametric model \citep{Yang2024}. Circles are colored by the maximum circular velocity of each Ms.Marvel halo at redshift zero. Black lines link collapse timescales from predictions and times measured directly from simulations for the same halo. The purple dashed line marks the time of the last simulation snapshot at redshift zero. 
  Right side: Predicted core-collapse timescales calculated using Equation~\ref{eq:parametric} (from \cite{Essig2019,Yang2023,Yang2024}) plotted against halo mass, using the same notation as the left side.}
  \label{fig:time_coll}
\end{figure*}

\subsection{Parameterization of gravothermal evolution}
\label{subsec:param}
\indent Following \cite{Nadler2024}, we analytically calculate the core-formation timescale, defined here as the full time that the SIDM halo is core-forming, ending at the point where the halo begins to undergo core-collapse, using a parametric model. The gravothermal evolution of each halo is parameterized by $\tau \equiv t/t_{cc}$ where $t$ is the formation time of each halo and $t_{cc}$ is the core-collapse timescale \citep{Essig2019}. Halos with $\tau < 0.15$ are defined as “core-forming” and halos with $\tau > 0.75$ are “core-collapsed". Halos that are core-forming have decreasing central densities and halos in the core-collapse phase have central densities exceeding CDM halos \citep{Outmezguine2023,Yang2023}. As such, the core-formation timescale is calculated as the time it takes for the halo to reach the point where $\tau = 0.75$ in its gravothermal evolution. This gravothermal evolution is governed by  a set of normalized density parameters such that when the halo becomes fully collapsed ($\tau = 1$) the scale density grows to be approximately 10 times its initial value \citep{Yang2023}. 
\newline\indent The formation time $t$ for each halo in Ms.Marvel is computed using a universal halo mass-formation time relation defined as when the NFW halo parameters $\rho_s$ and $r_s$  first become stable, estimated by its present day halo mass. See Section 3.1 of \cite{Yang2024} for more details. To analytically calculate the core-collapse timescales we use a parametric model \citep{Yang2023,Yang2024,Yang2025}, which models the gravothermal evolution of SIDM halos from their initial NFW profile. In this model the core-collapse timescale \citep{Essig2019} is characterized by the following equation;
\begin{equation}
  t_{cc} = \frac{150}{C} \frac{1}{(\sigma_{\rm eff}/m)r_{\rm eff} \rho_{\rm eff}} \frac{1}{\sqrt{4 \pi G \rho_{\rm eff}}}
  \label{eq:parametric}
\end{equation}
Where $\sigma_{\rm eff}$ is the effective cross section and $C$ is a constant, calibrated to N-body simulations. In this work, we adopt the same value of $C = 0.75$ used by \cite{Yang2024}.  We note that a larger (smaller) value of C would shorten (lengthen) the core-collapse timescales calculated in this work. We define the effective radius ($r_{\rm eff}$), and the density at the effective radius of the halo ($\rho_{\rm eff}$) below.
\begin{equation}
  r_{\rm eff} = \frac{R_\text{max}}{2.1626}, \qquad   \rho_{\rm eff} = \biggl(\frac{V_\text{max}}{1.648 \times r_{\rm eff}}\biggr)^2 \times G^{-1}
\end{equation}
In the case where the DM halo follows an NFW profile the effective density (radius) converges to the scale density (radius), see \cite{Yang2023} for more details. 
We calculate the effective cross section following work from \cite{YangYu2022} where we assume that DM particles follow a Maxwell-Boltzmann velocity distribution. The following equation assumes a differential cross section of $\frac{d\sigma}{d\cos\theta}$ to evaluate the cross section
\begin{equation}
    \sigma_{\rm eff} = \frac{1}{512 v_{eff}^8} \int dv d\cos\theta (\frac{d\sigma}{d\cos\theta}) v^7 \sin^2\theta  \exp(- \frac{v^2}{4v_{\rm eff}^2})
\end{equation}
where $v_{\rm eff} \approx 0.64$ V$_\text{max}$ (maximum circular velocity). The function integrates over the halo's characteristic velocity dispersion and scattering angle between DM particle collisions evaluated form from 0 to $\pi$. For a given halo, we assume a Maxwellian distribution and integrate the velocity term from zero to infinity. The relative velocity of DM particles is represented by $v$ and $\theta$ is the scattering angle in the center of mass frame. The differential cross section of SIDM particles in our simulations is approximated by the following equation from \cite{Tulin2013}
\begin{equation}
    \frac{d\sigma}{d\Omega} \approx \frac{\sigma_T}{4 \pi} \qquad \beta \lesssim1
\end{equation}
To apply this differential cross seciton to the equation described in \citep{Yang2023} we partially integrate the solid angle derivative over $d\phi$, assuming azimuthal symmetry. 
\newline\indent The parametric model predicts the gravothermal evolution of SIDM halos from the evolution history of the circular velocities of matched CDM counterparts. The model assumes that the mass accretion history of SIDM halos and CDM halos are similar. This assumption is primarily violated in subhalos, which are excluded from the halo sample used in this analysis. In this work, we study a sample of extremely low-mass SIDM halos from a cosmological simulation. Because of the stochastic nature of these simulations, particularly at such low masses, we only identify matched CDM halos for a few of the halos in our core-collapsed SIDM sample. As a result, we instead analytically calculate collapse timescales from the $V_\text{max}$ and $R_\text{max}$ of our SIDM halos at the final time-step in the simulation at redshift zero. In this case, the collapse time indicates how much additional evolution past redshift zero are required for the halo to become fully core-collapsed. Because we use the analytic collapse time to calculate the amount of time required for additional evolution we eliminate the underlying assumption that the mass accretion histories of SIDM and CDM halos are similar. 
\newline\indent However, this analysis is limited by the resolution of SIDM halos particularly at low masses and in the late stages of evolution. As we noted earlier, many of our halos may experience artificially elongated collapse times as a result of limited halo resolution. Additionally, it is extremely difficult to measure SIDM halos in the later stages of core-collapse. As a result we note that our analytically calculated collapse times, particularly at the low-mass end, are potentially biased towards longer collapse times.
\newline\indent Figure \ref{fig:time_coll} validates our measurement of $T_c$ measured from the inner density slopes of DM halos. Here we directly compare the core-formation and core-collapse timescales predicted by the parametric model \citep{Yang2023,Yang2024, Yang2025, Nadler2024}. The left side of Figure \ref{fig:time_coll} compares $T_c$ with the predicted core-formation timescales. The right side of Figure \ref{fig:time_coll} compares $T_c$ with the analytically predicted core-collapse timescale. The figure uses the following legend; blue squares mark $T_c$ for our sample of core-collapsed Ms.Marvel DMO halos and circles (diamonds) represent the core-formation (collapse) timescales and are colored by maximum circular velocity. Black lines link different timescales measured for the same halo. The dashed purple horizontal line marks the last timestep of the simulation, equal to the age of the Universe. 
\newline\indent Because the predicted core-collapse timescale is set such that it measures the amount of additional time past redshift zero required for the halo to become fully core-collapsed it is necessarily offset to longer timescales than the collapse timescale measured from inner density slope. We note that SIDM halos experience rapid gravothermal evolution after reaching the end of the core-forming phase, and should collapse on short timescales. As a result, the core-formation timescale can also be used to indicate the initialization of core-collapse in an SIDM halo.
\newline\indent We find that core-formation and core-collapse timescales are mass-dependent, where formation/collapse times increase with larger relative velocities between DM particles and decreasing cross-section of interaction. The predicted timescales of Ms.Marvel halos with masses greater than $5 \times 10^8 M_\odot$ grow steadily with increasing halo mass. Below this threshold, Ms.Marvel halos have a wide range of potential core-formation/collapse times. We note that this behavior is specific to the scattering model used in this work and different cross section models result in distinct patterns. However, at lower masses the gravothermal evolution of the halo is greatly influenced by halo concentration accross all SIDM models. As a result, for a halo to become core-collapsed within a Hubble time it must not only be low mass, but also have a high enough concentration to increase the rate of self-scattering events. 
\newline\indent The lowest mass Ms.Marvel halos have collapse timescales that range in several orders of magnitude. This huge scatter in core-collapse time that increases with decreasing halo mass may partially result from resolution effects \citep{Mace2024,Palubski2024}. \cite{Mace2024} found that for halos with fewer than $10^5$ particles the collapse rate has large variation due to the effects of large discreteness noise. Additionally, they found that these outliers in the collapse rate are slightly biased toward longer core-collapse timescales, which could partially account for the elongated collapse rates predicted in some of our analytic calculation in the UFD mass range, where halos may contain as few as $10^4$ particles. As a result, in halos with masses below $\sim 8\times10^8 M_\odot$ we would expect analytically predicted collapse times to be biased toward longer timescales. We note that the particle resolution effect is found to impact collapse times on an order of 10's of percent effect \citep{Mace2024}, indicating that some of the extremely long collapse times measured in this work may suffer from additional resolution effects. 
\newline\indent Nonetheless, in Figure \ref{fig:time_coll} we find good agreement between the collapse time directly measured from the inner density slopes of Ms.Marvel DMO halos $T_c$, and analytic calculations of core-formation and core-collapse. Despite only identifying 9 halos as core-collapsed out of the 24 halos with core-formation timescales shorter than the length of the simulation, we successfully find the halos with the shortest core-formation and core-collapse times. Each core-collapsed halo has a predicted core-formation time smaller than 10.5 Gyr. In contrast 9 out of the 24 halos with fully formed cores have core-formation times exceeding 12 Gyr. These halos require more than 13 Gyr of additional evolution past the end of the simulation to become fully core-collapsed. Compared to the halos we identify as core collapsed, seven are predicted to be fully collapsed within 2 Gyr of the last timestep, and all are predicted to be fully collapsed within 6 Gyr. We conclude that we are able to successfully identify the 9 most deeply core-collapsed halos in our sample. The remaining 15 out of 24 halos are likely at the onset of core-collapse, however, they are not yet collapsed enough to be detected. 
\newline\indent We find that the analytic core-formation timescales generally have good agreement with collapse times measured from the halo's inner density slope. All halos have $T_c$ that agree within 4 Gyr of the analytic prediction for core-formation, with no apparent differences in the degree of accuracy between lower and higher mass halos. Measured timescales appear to favor times that are slightly longer than the core-formation timescale, with only three halos (of the 9 that we identify as core-collapsed) having measured core-collapse times shorter than the analytically predicted core-formation time. This means that we identify halos as core-collapsed around or shortly after the halo has finished forming its core. These results indicate that even in the initial phases of core-collapse, the inner DM density profiles of SIDM halos become significantly cuspier. 
\newline\indent Most of the core-collapsed halos are all extremely low-mass ($M_{halo} \lesssim 2\times10^8M_\odot$) with central density cores that are too small to be fully resolved. We speculate that measuring the inner density slope closer to the center of these lower mass halos could bring our results into closer agreement. However, in this section we do not measure the density slopes inside of  0.33 kpc, the spatial resolution cut uniformly adopted in this work. In the Discussion section we study the effects of measuring the slope of core-collapsed halos at different radii, including measuring the density slope within three times the softening length (see Figure \ref{fig:core_rad}). For a discussion of spatial resolution limits adopted in this work see Appendix \ref{sec:app_B}. 
\newline\indent Previous work studying collapse timescales for SIDM halos \citep{Carton2022,zeng2025tillcorecollapsesevolution,Mace2024} commonly uses the increase or growth in central density as a metric to identify core-collapse timescales. In Appendix \ref{sec:central_den} we compare our results from this section with a different method for identifying halos in the core-collapse phase using measurements of the growth of central density of the halo. We find that, for identifying halos that are in the core-collapse phase, central density is a less accurate metric than inner density slope. For a more detailed discussion see Figure \ref{fig:den_comp} in Appendix \ref{sec:central_den} and the discussion therein. 

\section{Discussion}\label{sec:discussion}
\indent In this paper we study the gravothermal evolution of isolated dwarf halos simulated with SIDM by measuring the central densities and inner density slopes of halos from the Ms.Marvel simulation. Here we contextualize our results by comparing our simulations with other works and discussing the impact of measuring halo properties at different radii. We note that different papers favor a variety of methodologies for measuring density slope and both observers and simulators frequently measure halo properties at different radii, complicating the comparison of our results with previous works. In the following section we discuss how differences between simulations, such as SIDM model and environment, can impact the gravothermal evolution of SIDM halos. We also measure halo properties across a range of radii, these results have important implications for how the definition of core-collapse changes, depending on the radius of measurement.  

\subsection{A comparison to other SIDM simulations}
\begin{figure*}
  \centering
  \includegraphics[width=1\linewidth]{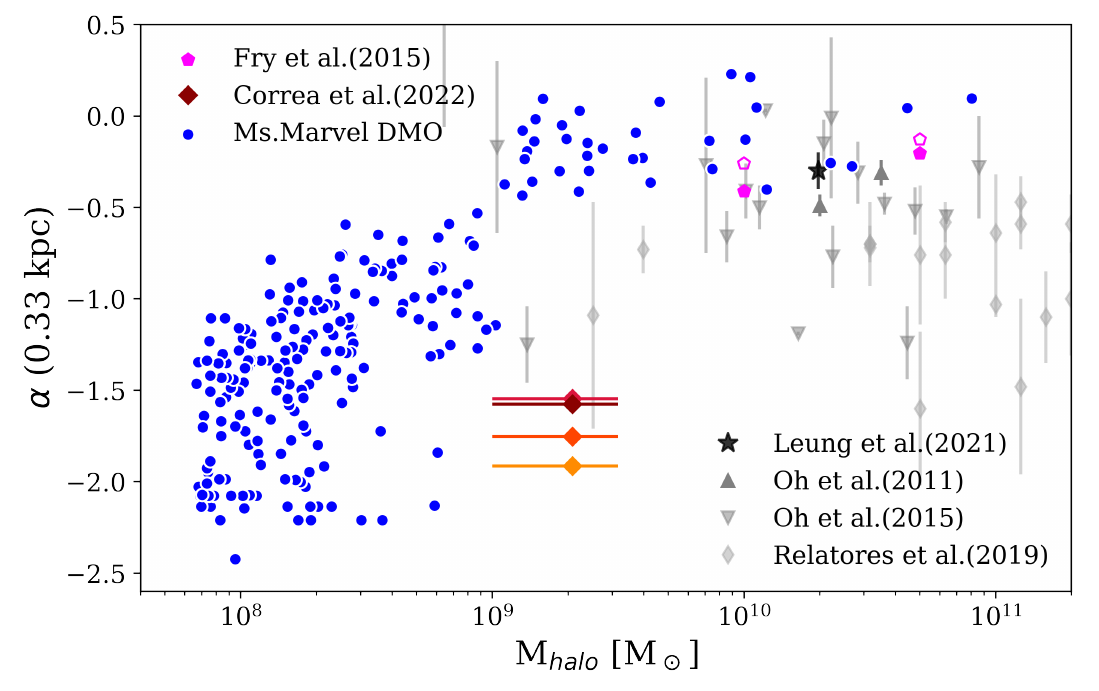}
  \caption{Central inner density slope ($\alpha$) vs DM halo mass for SIDM halos from the Ms.Marvel DMO run (blue circles), \cite{Correa_2022} (colored diamonds), and \cite{Fry2015} (pink pentagons). Open points are DMO runs while filled points are baryonic runs.   Grey points represent observed slopes in dwarf galaxies from \cite{Leung2021} (square), \cite{Oh2011slopes} (up triangle), \cite{Oh2015} (down triangle), and \cite{Relatores2019} (diamond). Inner density slope is calculated at 0.33 kpc from the DM density profile. }
  \label{fig:sigma_comp}
\end{figure*}

\indent In the following section we compare we compare our Ms.Marvel DMO halos with different SIDM cross sections from \cite{Fry2015} and \cite{Correa_2022}. In Section \ref{sec:sim} we compare the cross sections from this work with the cross sections from \cite{Correa_2022} and \cite{Fry2015} directly (Figure \ref{fig:sigma_eff}), however, Figure \ref{fig:sigma_comp} plots the inner density slopes of of SIDM halos from various models against their halo mass. Blue points represent Ms.Marvel halos, where inner density slope is calculated using the same method described in Figure \ref{fig:Mvalpha}. Grey markers represent the density slopes of observed galaxies as described in Figure \ref{fig:Mvalpha}. Pink hexagons represent halos from \cite{Fry2015} with a constant cross section of interaction of $\sigma = 2$ cm$^2$/g. They measure the DM density profile measured at 500 pc for two distinct field dwarfs with masses of $10^{10} M_\odot$ and $5 \times 10^{10} M_\odot$. We find that, for halos in this mass range, the cross section implemented by \cite{Fry2015} is consistent with the cross section adopted in this work. This is, perhaps, unsurprising considering that at the effective cross section of our velocity-dependent SIDM model lies between $10^{0}$ and $10^{1}$ for halos in this mass range. 
\newline\indent The results presented from \cite{Correa_2022} are the average slopes of a sample of satellites of MW-mass hosts from the TangoSIDM simulations with masses ranging from $10^{9-9.5} M_\odot$. \cite{Correa_2022} measure the logarithmic density slope by taking a linear fit of the DM density profile between $1-3$ kpc. The different colors correspond to four unique SIDM models; SigmaConstant10 (dark red), SigmaVel20 (red), SigmaVel60 (dark orange), and SigmaVel100 (lighter orange). SigmaConstant10 is a constant cross section of interaction $\sigma = 10$ cm$^2$/g, while the velocity-dependent models are characterized by the cross section reached at 10km/s such that, for SigmaVel100 $\sigma_T/m_\chi = 100$ cm$^2$/g. Similarly models reaching 60 and 20 cm$^2$/g at 10 km/s are called SigmaVel60 and SigmaVel20, respectively. Figure \ref{fig:sigma_eff} plots the transfer cross sections of out SIDM model and the models from \cite{Fry2015} and \cite{Correa_2022} for a direct comparison. We show in Figure \ref{fig:Mvalpha} that our SIDM model allows us to match observations of central cores in isolated field dwarf galaxies in the bright dwarf mass range ($M_{halo} = 10^{10-11}M_\odot$). At lower masses, a subset of SIDM halos undergo core-collapse, becoming extremely centrally dense to match low-mass objects with high central densities \citep{Oh2015,Relatores2019}. 
\newline\indent Similarly to the SIDM model studied in this work, \cite{Correa_2022} also simulate a light-mediator model with a Yukawa potential. However, the differential cross section from the Ms.Marvel simulations lies in the perturbative regime (where $\alpha_\chi m_\chi / m_\phi < 1$), while the Transfer cross section in this work considers non-perturbative corrections outside of the regime where the Born approximation is valid. 
\newline\indent The cross sections of the SIDM models implemented by \cite{Correa_2022} are lower than the velocity-dependent interaction cross section implemented in Ms.Marvel, however, we find that TangosSIDM halos undergo gravothermal collapse at higher masses relative to Ms.Marvel DMO halos. We expect that differences in environment may account for the accelerated collapse at higher masses in the TangoSIDM simulations. TangoSIDM halos are satellites of a MW-mass host while the largest halo simulated in Ms.Marvel is a Large Magellanic Cloud (LMC) mass halo. Additionally, we exclude satellites in our sample, meaning that all Ms.Marvel halos presented in this work exist in a relatively isolated environment. This speculation is supported by \cite{Nadler2023,Yang_2023} who show that core-formation in SIDM halos is dependent on environment. \cite{Nadler2025} find that the rate of core-collapse is also higher in subhalos relative to halos simulated in isolation. Additionally, previous work shows that the gravothermal evolution of SIDM halos can be accelerated or disrupted by tidal interactions \citep{Carton2022,Jiang19,Benavides2021,Benavides2023}. These results underscore the importance of studying core-formation in SIDM across a range of environments. 

\subsection{Tracing inner density slope at different radii}
\label{subsec:aperture}
\begin{figure*}
  \centering
  \includegraphics[width=1\linewidth]{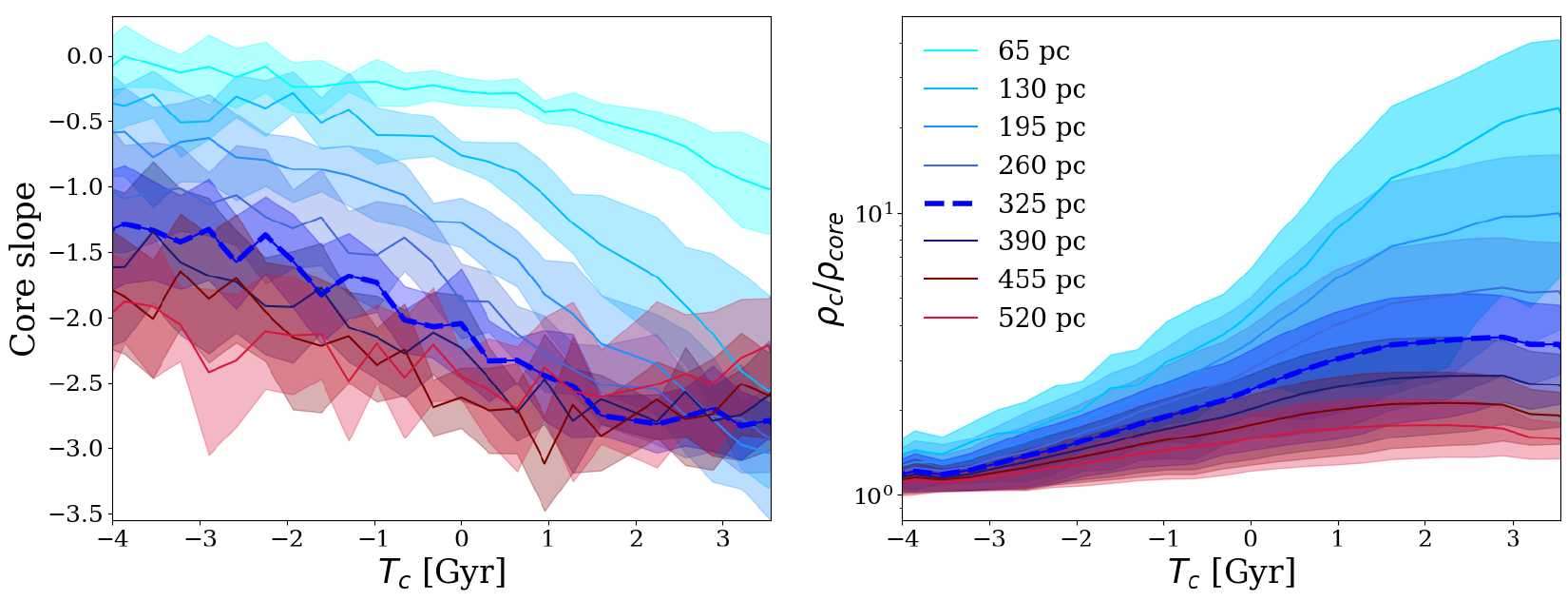}
  \caption{Left side: inner density slope vs. $T_{c}$ averaged for nine core-collapsed halos at different radii ranging from one softening length at 65 pc (light blue) to 8 times the softening length at 520 pc (red). The radius of measurement is colored using the same legend shown on the right-hand side for both inner density slope and central density. Lines represent the average inner slope traced through time at each radius for the entire sample of core-collapsed Ms.Marvel DMO halos with the shaded region corresponding to one standard deviation. $T_{c}$ is the time relative to core-collapse such that each halo's collapse time is defined as $T_{c} = 0$. Core-collapse time is calculated using the same methodology as in Figure \ref{fig:trace_core}.
  Right side: Central density normalized by the density of the fully formed core vs. $T_{c}$. Central densities are measured at the same radii and use the same legend and x-axis as in the left-hand side.}
  \label{fig:core_rad}
\end{figure*}
 \indent As discussed in the previous section, the DM density profiles of Storm CDM DMO, Storm CDM+baryons, and Ms.Marvel DMO halos are not fully resolved at central radii below 0.33 kpc. This resolution limit severely restricts our ability to study the centers of the lowest-mass halos in our sample. Both Figures \ref{fig:Mvalpha} and \ref{fig:sidm_trace_alpha} demonstrate that the inner density slopes of Ms.Marvel DMO halos exhibit a strong mass dependence, where slope decreases with the halo mass. We find that low-mass halos (M$_{halo} < 10^9 M_\odot$) form cores with sizes well below the resolution limits of the simulations. As a result, inner density slope measurements for all halos below this mass threshold take the slope of the profile well outside the core radius. 
 \newline\indent These results indicate that fully resolving core-formation in UFD-mass halos requires extremely high resolution, which would be difficult to achieve in cosmological simulations, especially in simulations that incorporate baryonic processes. However, we can still obtain information about the centers of these low-mass halos even without fully resolving their cores. Our results show clear evidence of core-collapse in halos with halo masses below $2\times10^9 M_\odot$. Of these nine, the halos with the highest central densities have inner density slopes approaching $-3$, far below those predicted by CDM simulations as seen in Figure \ref{fig:Mvalpha}. These slopes also deviate from average inner density slope values in their given halo mass bin. We conclude that the process of core-collapse impacts the shape of the DM density profile out to much larger radii than the core radius. Hence, profile shapes outside of the central DM density core can still give us insight about the behavior of DM at halo centers. 
\newline\indent To further test how the evolution of the density profiles of core-collapsed halos is impacted by spatial resolution constraints, we measure inner density slope at different radii for our sample of nine core-collapsed Ms.Marvel DMO halos. The left side of Figure \ref{fig:core_rad} plots the average density slope measured at different radii, marked by color, over time. We measure inner density slopes at 7 different radii ranging from 65 pc (blue) to nearly 500 pc (red) for core-collapsed Ms.Marvel DMO halos. Time is measured with respect to core-collapse time, where core-collapse time is defined by inner density slope as in Figure \ref{fig:time_coll} with the time of core-collapse ($T_{c}$) set equal to zero. We define time prior to the onset of core-collapse as negative while positive times represent the number of years after core-collapse. Shaded regions represent one standard deviation from the mean. 
\newline\indent We choose to study radii ranging from 65 pc to roughly 500 pc to represent a broad scale of different spatial resolution limits across simulations and observational data. The smallest limit we consider here is the gravitational softening length, while the largest radius is large enough to be directly compared with observations. As shown in Appendix \ref{sec:app_A}, the growth of central density in core-collapsed halos remains stable up to approximately the softening length. As a result, we believe that while the regions we study are unresolved within 0.33 kpc in accordance with stringent resolution restrictions, we expect these trends to be largely consistent at higher resolution. Furthermore, without extensive resolution tests trained on SIDM halos it is unknown if SIDM simulations face the same resolution issues documented in CDM. 
\newline\indent We find that measuring the inner density slope closer to the halo center results in values that are more cored. Additionally, we find that at all radii the slopes of Ms.Marvel DMO halos are shallowest several Gyr before they begin to collapse. Within one softening length, our subsample of Ms.Marvel DMO halos remain cored until they enter the initial stages of core-collapse. In contrast, slopes measured outside of this radius steepen during the core-forming phase before the halos begin undergoing collapse. The inner density slopes measured at radii outside of three softening lengths (180 pc) converge below $-2.0$ within a couple of Gyr after the onset of core-collapse. Inside of three times the softening length we find that slopes remain shallower for more extended periods of time, potentially resulting from lowered central densities in unresolved halo centers.
\newline\indent At radii outside of 400 pc we find that inner density slopes start growing shallower approximately 1 Gyr after core-collapse. It is reasonable to guess that this behavior might be related to the turndown effect that we identify in Figure \ref{fig:trace_core} and discuss in Appendix \ref{sec:app_A}. However, we show in Figure \ref{fig:trace_core} that the inner slopes of these halos measured at 325 pc don't grow shallower as the central density decreases. This indicates that the slope of the DM density profile evolves differently at radii inside and outside of the central DM density core. We recommend measuring the slope of the DM density profile at the halo's core radius. In the case that the core radius lies within the resolution limits of the simulation we recommend measuring the slope of the profile at the smallest possible resolved radius. 
\newline\indent While our results indicate that the inner density slopes of core-collapsed SIDM halos are cusped enough to be distinct from the inner density slopes of CDM halos at a variety of radii, it is still difficult to observationally identify core-collapsed halos. This is largely because UFDs are extremely low-mass systems, making it hard to resolve their central DM density slopes. For instance, \cite{McConnachie12} find that UFDs can have half-light radii ranging from tens of parsecs to several hundred parsecs for galaxies with $M_* < 10^5 M_\odot$ in the Local Group. By measuring the slope of the DM density profile at larger radii we may be able to probe the inner slopes of DM halos at lower masses. We find that even though the measured density slope of Ms.Marvel DMO halos in the initial stages of core-collapse depend on the radius of measurement, our results converge outside of three times the softening length. This is in agreement with previous tests that indicate a convergence radius of 2.8$\epsilon$ \citep{Ludlow19}. We note that 2.8$\epsilon$ is well tested for CDM, but we may be able to resolve lower masses in SIDM simulations because SIDM scatter dominates over the spurious two-body gravational interactions. In Appendix C we briefly discuss how resolution criteria trained on CDM simulations do not account for SIDM self-scatterings and may not be applicable to SIDM simulations. As a result, inner density slope remains a valid mechanism for identifying core-collapse out to 500 pc.

\subsection{Tracing central density at different radii}
\indent We compare the effect of measuring inner density slope at different radii with central density in the right side of Figure \ref{fig:core_rad} by measuring central densities at different radii. Here we plot the normalized density measured at different radii separated by color over time. Central densities are normalized by dividing by the central density of the fully formed core for each halo. We plot the mean of this value for the nine core-collapsing halos over time. 
\newline\indent The right side of Figure \ref{fig:core_rad} demonstrates that the central densities measured at different radii diverge as the halo enters the core-collapse stage. When measuring central density at radii exceeding 390 pc the average halo only grows to at most twice as centrally dense as its initial central density. In contrast, measuring central density at 65 pc results in the central density increasing by more than an order of magnitude at the late stage of core-collapse. The changes in the growth of central density between different radii are most stark in the more advanced stages of core-collapse, however, these differences are apparent even at the initialization of core-collapse. Central density grows rapidly within the innermost regions of the halo and grows at a slower rate in the outer regions of the halo. 
\newline\indent The changes in the rate of growth in the central density of core-collapsed halos at different radii has important implications for identifying core-collapsed halos observationally and constraining core-collapse timescales. Firstly, these results indicate that it may be very difficult to observationally distinguish collapsed SIDM halos from their cored counterparts in the UFD mass range by their central densities. Secondly, when implementing a threshold for collapse, it is important to consider at what radii halo properties are measured. We find that central density, in particular, is extremely sensitive to radius compared to the slope of the DM density profile. We conclude that while central density remains a more direct indicator of core-collapse in well-resolved SIDM halos, measuring the inner density slope is more accurate in lower resolution simulations and at larger radii of simulated halos. 

\begin{figure*}
\centering
\includegraphics[width=1\linewidth]{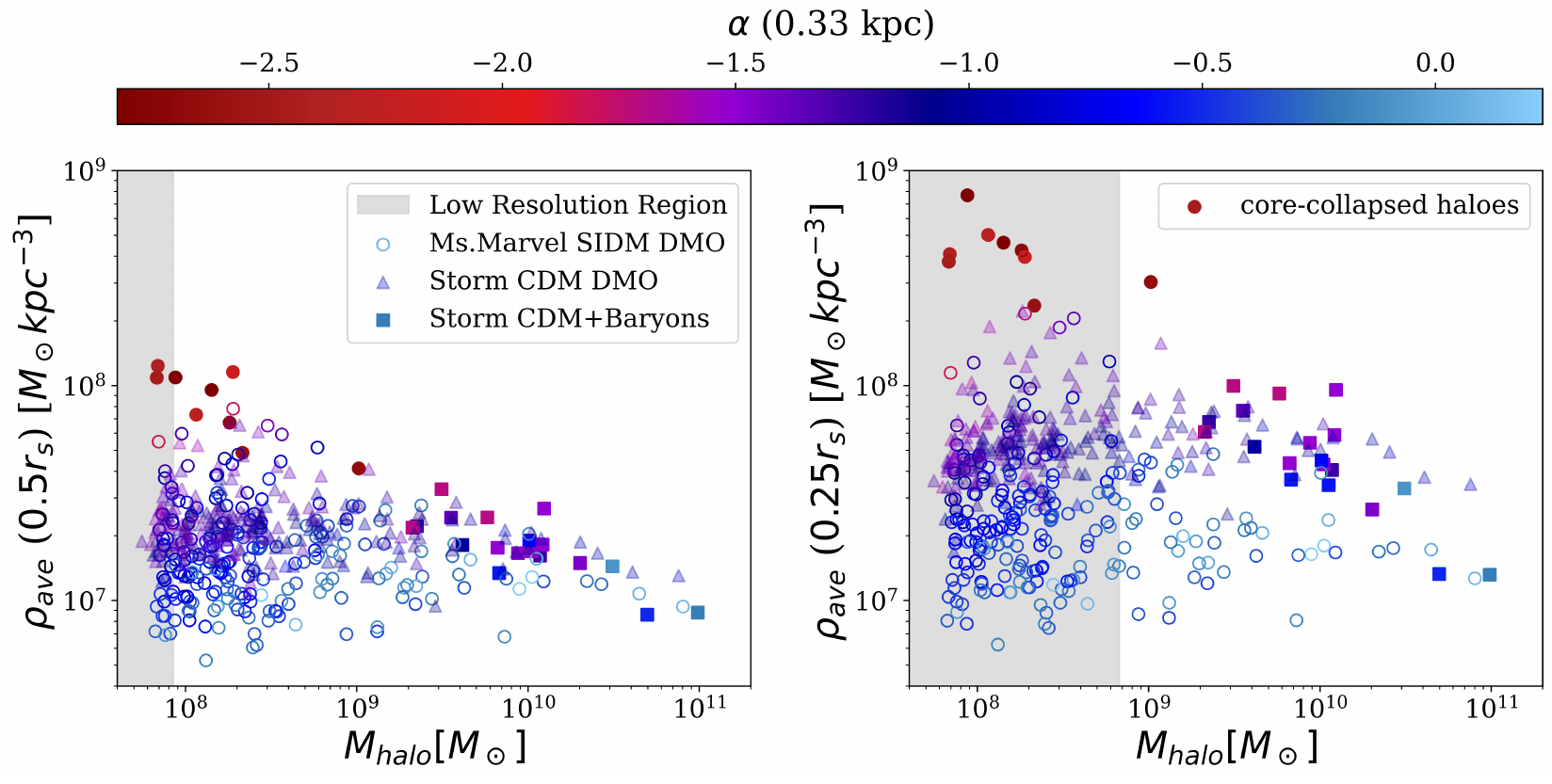}
\caption{Left side: the averaged density contained within $50\%$ of a characteristic scale radius at redshift zero for Ms.Marvel DMO halos (empty circles), Storm DMO halos (filled triangles), and the baryonic run of Storm (filled squares). Each point is colored by the inner dark matter density slope, measured using the same methodology as in Figure \ref{fig:Mvalpha}. The filled circles represent SIDM halos that we identify as core-collapsed using the criteria from Figure \ref{fig:time_coll}. The grey shaded area represents a ``low resolution" region where for low-mass halos the $50\%$ and $25\%$ fractions of the scale radius are smaller than the spatial resolution of the simulation (0.33 kpc). 
Right side: this figure uses the same shared notation from the left-hand side but plots the central density averaged within  $25\%$ of the characteristic scale radius.}
\label{fig:scale_rad}
\end{figure*}
\subsection{Potential kinematic signatures of core-collapse}
In Section \ref{subsec:aperture}, we discuss the viability of identifying core-collapsed halos observationally from measurements of their central densities and the inner slopes of their dark matter density profiles.  However, stellar kinematics may also be used to potentially identify core-collapsing SIDM halos. \cite{zeng2025diversityuniversalityevolutiondwarf} find that in idealized simulations of a halo with $M_{200} = 3\times10^{10} M_\odot$ (where $M_{200}$ is the mass of a halo with a radius that is 200 times the critical density) both the half-light radius and the mass-to-light ratios within the half-light radius were significantly altered by core-collapse. They find that the half-light radius of core-collapsed halos could shrink down to half their original radius, while the mass-to-light ratios increased by an order of magnitude. We anticipate that hydrodynamic runs of the core-collapsing halos in our sample, which have central densities that increase by an order of magnitude within $\sim 100 - 200$ pc of the halo center,  should experience a similarly detectable growth in mass-to-light ratio.
\indent Figure \ref{fig:scale_rad} plots the averaged central density ($\rho_{ave}$) contained within  $50\%$ and $25\%$ of a characteristic scale radius ($r_s$) for each halo in the Ms.Marvel SIDM DMO, Storm CDM DMO, and Storm CDM+baryon simulations. The characteristic scale radius for each halo is determined by a best fit to the linear relationship between halo mass and the scale radius of CDM DMO halos from the Storm simulations, fit with an NFW profile \citep{Navarro96} in log-log space. We find that for Storm DMO halos this relationship contains little scatter, making halo mass an excellent predictor of the value of the scale radius for an unaltered NFW profile. We note that the smallest characteristic scale radius in our sample is equal to 0.65 kpc and, as a result, the measurements of the averaged central density for each halo at half the scale radius for all but the lowest mass halos in our sample remain well-resolved. We highlight the regions of low resolution with grey shaded areas. We caution that our results on the right panel measure the central density of each halo within up to $\approx$0.16 kpc, and note that halos with masses $M_{halo} \lesssim 7 \times 10^8 M_\odot$ lie within a region not considered fully resolved by our conservative estimates. Therefore, the central densities of the lowest-mass halos in our sample should be taken as a lower bound.
\newline\indent We find that core-collapsing SIDM halos, highlighted as the filled circles while the core-forming SIDM halos are marked by empty circles, are more centrally dense than their CDM and core-forming counterparts. Measuring at a smaller fraction of the scale radius increases the difference between core-collapsing and core-forming SIDM halos, with all core-collapsing halos having higher central densities than any other halo in our simulations. The core-collapsed halos reach central densities within $25\%$ of the scale radius that are up to ten times higher than their CDM counterparts, which marks central density as a promising indicator of core-collapse. However, it may be difficult to accurately measure central densities within the radii considered here, which are typically on the order of one hundred to a few hundred parsecs in halos less massive than $M_{halo} = 10^9 M_\odot$. Interestingly, the highest-mass SIDM halos appear more centrally dense than the highest-mass CDM halos from the baryonic Storm run, though both have similar inner DM density slopes.

\subsection{Tidal effects in SIDM subhalos}
\indent In this work, we analyze halos simulated in a low-density environment, excluding satellite dwarfs from the study. We note that most observed dwarf galaxies, especially in the ultra-faint mass range, are satellites in the Local Group. While an in depth study of the impact of tidal effects on the gravothermal evolution of SIDM halos is beyond the scope of this work, we provide a brief overview here.
\newline\indent Previous works show that SIDM can suppress the abundance of subhalos through enhanced tidal stripping and tidal disruption of cored halos and evaporation from subhalo–host halo interactions \citep{Vogelsberger2012,Nadler20,Errani_2022}. High resolution cosmological simulations of MW-mass halos and their environment find that SIDM suppresses the subhalo mass function by up to $50 \%$ relative to CDM \citep{Nadler2025}. Similarly, for a cross section of  $\sigma / m_\chi = 2 ~\rm{cm^2g^{-1}}$,  \cite{Nadler20} find that $56 \%$ of subhalos were disrupted and erased due to subhalo-host halo interactions. However, \cite{Robles_2019} find that, for $\sigma / m_\chi = 1 ~\rm{cm^2g^{-1}}$, the predicted substructure abundance of MW satellites is very similar between SIDM and CDM. Surveys probing the abundance and structure of the MWs satellite population will provide important constraints for the cross section of SIDM \citep{Nadler2025,Correa2021}. 
\newline\indent The interaction cross section of SIDM models may also be constrained by the density profiles of dark matter halos. Velocity-dependent models with high interacting cross sections that induce core-collapse at low masses predict a wide diversity of density profiles. These SIDM models produce halos that may explain observations from the very low-concentration ultra-diffuse galaxies \citep{zhang2024gd1stellarstreamperturber,Roberts_2025} to extremely concentrated substructures detected by gravitational imaging \citep{YangYu2022,Nadler2023,zeng2025tillcorecollapsesevolution}. \cite{Nadler2024} find that, for an SIDM model with strong velocity-dependence, high mass SIDM subhalos are cored, while subhalos with $M_{halo} \leq 10^9 M_\odot$ have density profiles that range from extremely cored to very cuspy for core-collapsed objects. Additionally, they find that cored SIDM halos exist in a much higher abundance in isolation compared to satellites. Cored subhalos are more likely to become tidally disrupted \citep{Errani_2022}, but the decreased number of cored subhalos is also likely impacted by other factors. The gravothermal evolution of SIDM subhalos has been shown to be accelerated by tidal stripping \citep{Nishikawa2020, Sameie2020, Carton2022}. Future hydrodynamic SIDM simulations will test whether these models are capable of reproducing the tight scaling relations exhibited by dwarf galaxies in the Local Group.

\subsection{Halo mergers and substructure in SIDM halos}
\indent It is predicted that, even at lower masses, dark matter halos may contain additional substructure in the form of even lower mass subhalos. We exclude all satellites from our analysis, focusing on halos that are located in relative isolation and we note that subhalos of most of our sample would lie below the resolution of the simulation and remain unresolved. As a result, we do not directly test how our results may be impacted by halo substructure, however, we briefly discuss it here.
\newline\indent In Section \ref{subsec:trace} we find that in the 1-2 Gyr immediately following a merger event the central densities and inner density slopes of SIDM halos can fluctuate significantly. In Figure \ref{fig:sidm_trace_alpha} we found cases where the inner slopes of the DM density profile temporarily flattened and at other times where the inner slope became steeper after a merger. Additionally, we find evidence in Figure \ref{fig:time_coll} that major mergers may delay the onset of core-collapse in SIDM halos. These results imply that mergers can temporarily alter the DM density profiles of DM halos. Recent work from \cite{silverman2026mergersmattergravothermalcollapse} finds that the gravothermal evolution of SIDM halos is altered by halo mergers, where halos with sustained mergers never undergo core-collapse.
The impact of halo mergers on the density profiles of halos may also be more pronounced in SIDM than in CDM. \cite{Kim_2017} find that SIDM halos contain a higher portion of gravitationally unbound particles and that it is easier to unbind particles from SIDM mergers due to the shallower potentials of cored SIDM halos relative to cuspy CDM halos.
\newline\indent Recently, \cite{penarrubia2025gravothermal} found that dark subhalos can produce fluctuations to the gravitational potential of the host halo resulting in dynamical heating of the central regions and a gradual expansion over time. This effect is most pronounced in low-mass dwarf galaxies. These results assume that dark matter is a collisionless particle. With SIDM halos it is unclear how much infalling dark halos would impact the central potential well of the host and it's internal dark matter structure. There are two main differences that could significantly impact this process in SIDM. The first is tidal disruption of subhalos: cored SIDM halos are more susceptible to tidal disruption than their CDM counterparts \citep{Vogelsberger2012,Nadler20,Errani_2022}. If the subhalos become fully disrupted, could their dark matter particles still dynamically heat the center of the host halo effectively? The second is that the gravitational potential wells of cored SIDM halos are shallower to begin with, which makes them potentially more susceptible to additional disturbance by substructures \citep{Kim_2017}. 
In this work, we show that major disruptions to the centers of SIDM halos can delay the onset of core-collapse, however, disturbances from tidal stripping have been shown to accelerate the gravothermal evolution of SIDM subhalos \citep{Nishikawa2020, Sameie2020, Carton2022}. Additionally, \cite{kong2026gravothermalcollapserobustbaryonic} find that strong stellar feedback models can delay the onset of core-collapse in high concentration halos, while weak feedback mildly accelerates core-collapse. \citep{kong2026gravothermalcollapserobustbaryonic} also find that, for median concentration SIDM halos, stellar feedback at early times can rapidly produce large, shallow cores during the core-formation phase. As a result, we expect that the injection of energy through the accretion of dark subhalos at early times might similarly accelerate core-formation in SIDM halos and lead to shorter core-collapse timescales. We note that this process could also delay the onset of core-collapse in halos that have low concentrations, or where the energy injected by subhalo accretion is sufficiently large.

\section{Conclusions}
\label{sec:conclusions}
In this work we explore the distribution of DM at the centers of DM halos from three high-resolution, zoom-in simulations run with the same initial conditions: Ms.Marvel DMO (SIDM), Storm CDM DMO (CDM), and a hydrodynamic run of Storm (CDM +baryons). Our SIDM model implements a velocity-dependent cross section of interaction which preserves the predictions of CDM at large scales but naturally forms cores at lower masses, reducing tensions with the ``cusp-core" problem. We analyze the inner density slopes and densities at the centers of DM halos at redshift zero, contextualizing our results with observations. By tracing these properties through time, we track the evolution of CDM and SIDM halos and find evidence of core-collapse in 9 Ms.Marvel DMO halos. We use a parametric model of gravothermal evolution to analytically calculate core-formation and core-collapse timescales of our Ms.Marvel DMO halos, showing that our identification of core-collapsed halos is reasonable. Using these analytic calculations we test different methods of identifying core-collapsed SIDM halos and explore their dependence on radius of measurement. Our results can be summarized as follows:
\begin{itemize}

\item Between halo masses of $10^{9-10} M_\odot$, predictions from Storm CDM+baryons and Ms.Marvel DMO diverge, indicating that we may be able to distinguish between DM models at these scales observationally (Figure \ref{fig:Mvalpha}). We find that SN feedback is only effective at creating cores in the regime of strong feedback. In contrast, SIDM models can tune the cross section of interaction such that SIDM halos are core-forming at lower masses relative to current CDM+baryonic simulations. Core-collapsing halos have inner DM density slopes that are steeper than $-1.9$, the lowest slope we measure in any Storm CDM DMO halo. Observations of such centrally dense DM halos favor SIDM models with high velocity-dependence and may help to rule out CDM. 
\item Core-formation takes place on shorter timescales in Ms.Marvel SIDM halos compared to Storm CDM halos with baryonic feedback (Figure \ref{fig:sidm_trace_alpha}). In our core-collapsed sample of halos from the Ms.Marvel DMO simulation, cores initially form during the first 6-8 Gyr of the halo's lifetime, and then inner density slopes become steeper at later times due to the onset of gravothermal collapse. In contrast, Storm CDM DMO halos are generally cuspy throughout their lifetimes (Figure \ref{fig:cdm_trace_alpha}). 
\item We find that central density traces inner density slope in Ms.Marvel DMO halos (Figures \ref{fig:sidm_trace_alpha} and \ref{fig:sidm_trace_den}) even in the case where the halo has initialized the core-collapse phase as shown in Figure \ref{fig:trace_core}. As the cores of Ms.Marvel DMO halos begin collapsing, their central densities grow to be several times denser and they have increasingly negative inner density slopes, exceeding those of their Storm CDM DMO counterparts. 
\item Figure \ref{fig:time_coll} indicates that, even in the initial stages of core-collapse, SIDM halos can potentially have inner density slopes steep enough to be distinct from CDM.  
\item  At radii of 0.5 kpc, which are large enough to be resolved by observers, we find that the central densities in core-collapsed halos do not grow enough to be identifiable as having core-collapsed  (Figure \ref{fig:core_rad}). Additionally, we find that inner density slope changes less with an increasing radius of measurement (Figure \ref{fig:core_rad}) and is a more accurate tracer of core-collapse than central density (Figure \ref{fig:time_coll}). As a result, it may be easier to identify core-collapsed halos observationally by the slope of their DM density profiles. 
\item We find that it is possible to characterize core-collapse even when the halo's core is goes below than the traditional resolution criterion. The radii of central DM density cores in all but the most massive halos from our sample of Ms.Marvel DMO halos are generally smaller than three times the gravitational softening length. However, the inner density slopes and densities measured at or just above our spatial resolution threshold of 0.33 kpc still contain information about the behavior of DM at the very center of SIDM halos. 
\end{itemize}

\subsection{Future work}
\indent Comparing observations with hydrodynamic runs of SIDM simulations will provide additional insight to the particle nature of DM. As hydrodynamic runs of CDM simulations have shown, incorporating baryonic feedback can drastically impact the distribution of DM at the centers of halos, particularly those that host bright dwarf galaxies. We will robustly test our velocity-dependent SIDM model in future runs with SIDM and baryons. Recent work indicates that the inclusion of baryonic feedback may not yield significant differences in the measured inner density slopes of Ms.Marvel DMO halos. \cite{Vargya2022,Zentner2022} show that the inner structure of SIDM halos is not as sensitive to baryonic physics as CDM for halos ranging in mass between roughly $10^{10-13} M_\odot$. Additionally, baryonic feedback processes take place on timescales larger than the evolution timescale of isothermal cores formed in SIDM halos, which allows core-formation to be dominated by the gravothermal evolution of SIDM halos. However, \cite{Straight_2025} find that baryonic processes do contribute a small amount to establishing the density profiles of SIDM halos. Future hydrodynamic runs of Ms.Marvel will test the sensitivity of core formation in SIDM halos to baryonic processes as well as their impact on core-collapse timescales. 
\newline\indent It is important to understand the role of baryons in SIDM simulations; for example, recent work shows that including baryons can make cores smaller, where these effects are most pronounced in baryon dominated galaxies \citep{kaplinghat2014,Robertson2019,Rahimi2023}. \cite{Robles2017} find that SIDM is more robust to the inclusion of baryons than CDM in dwarf galaxies ($M_* \approx 10^{5.7-7.0} M_\odot$), meaning that it may be more difficult to constrain uncertain baryonic physics in SIDM simulations. This added difficulty indicates that SIDM simulations may also reduce observable differences that arise from variations in subgrid modeling. A full understanding of baryonic effects on the gravothermal evolution of SIDM halos will be necessary to place constraints on the SIDM cross section. Future work will focus on simulating SIDM halos with the inclusion of baryons. 
\newline\indent The predictions made in this work may play a key part in constraining DM models with upcoming and current surveys. The timescales of gravothermal core-collapse in SIDM halos may also be used in future efforts to constrain the velocity-dependence of the cross section of interaction. The velocity-dependence of the interacting cross section for SIDM particles can be tuned such that only dwarf halos below a specific mass threshold will undergo core-collapse by redshift zero. However, we find that the varied merger histories of SIDM halos may impact their gravothermal evolution, with major mergers potentially delaying the onset of core-collapse. Some previous work places constraints on the velocity-dependence of SIDM using the density profiles of MW dwarf spheroidal galaxies \citep{Read17,Valli_2018,Correa2021}. As discussed previously, environment may impact the gravothermal collapse of SIDM halos \citep{Jiang19,Benavides2021,Carton2022,Benavides2023,Nadler2023}, indicating that it is necessary to also study the density profiles of isolated dwarfs to constrain the velocity-dependence of SIDM. 

\appendix 
\section{Using central density as a measurement of core-collapse}\label{sec:central_den}
\begin{figure*}
\centering
  \includegraphics[width=1\linewidth]{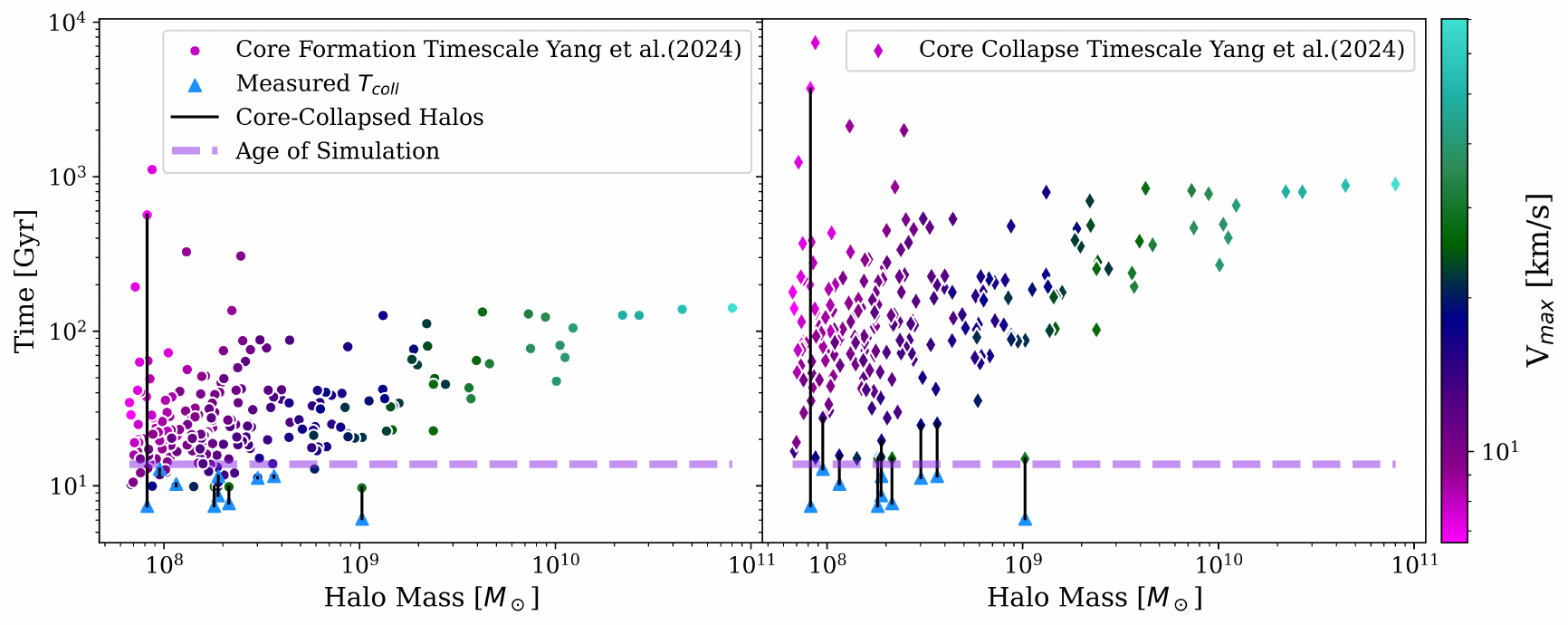}
  \caption{Left side: Predicted core-formation timescales plotted against halo mass for our sample of core-collapsed Ms.Marvel DMO halos. Blue triangles mark the time when the central density of a Ms.Marvel DMO halo first reaches 1.5 times its minimum central density averaged within 0.35 kpc, while colored circles represent the predicted core-formation timescales, calculated analytically from a parametric model. Circles are colored by the maximum circular velocity of each halo. Black lines link collapse timescales from predictions and times measured directly from simulations for the same halo. The purple dashed line marks one Hubble time. 
  Right Side: Predicted core-collapse timescales plotted against halo mass, using the same notation as the left side.}
  \label{fig:den_comp}
\end{figure*}

\indent Previous work studying collapse timescales for SIDM halos \citep{Carton2022,zeng2025tillcorecollapsesevolution,Mace2024} commonly uses the increase or growth in central density as a metric to identify core-collapse timescales. We compare our results from this section with a different method for identifying halos in the core-collapse phase using measurements of the growth of central density of the halo. Figure \ref{fig:den_comp} recreates results from Figure \ref{fig:time_coll}, but now compares analytic predictions with a collapse time based on the growth in central density. Halos are defined as core-collapsed if they have central densities which are 1.5 times their minimum central density averaged within 0.35 kpc. We measure the minimum central density of the halo once it has formed a core, making it distinct from the halo's initial central density. We define $T_{coll}$ as the time it takes for a halo to reach this threshold for growth in central density. We find that 4 of the lowest mass halos with core-formation timescales shorter than a Hubble time are not identified successfully. Additionally, this method falsely identified two halos with collapse times far exceeding a Hubble time. One of these halos has a late-time merger which accounts for the growth halo in halo density, indicating that this metric is not robust to halo mergers. 

\indent In the Discussion we show that central density is extremely sensitive to the radius at which it was measured, so accuracy will be improved with higher resolution or by eliminating lower-mass halos from the sample. 
\newline\indent Figure \ref{fig:den_comp} compares the analytically predicted core-formation and core-collapse timescales of Ms.Marvel halos against a core-collapse timescale identified by measuring the growth in central density over time. We identify 10 Ms.Marvel halos with central densities that grow by a factor of 1.5 over time, including two intermediate-mass halos ($M_{halo}\sim3\times10^8 M_\odot$) that finish forming their cores at late times and are not predicted to fully collapse until roughly 15 Gyr past the last timestep in the simulation. Additionally, we identified the 7 halos with the shortest core-collapse times with halo masses greater than $\sim 2 \times 10^8 M_\odot$ as core-collapsing. This indicates that central density is sensitive to core-collapse in halos that are sufficiently massive. However, at lower masses central density measurements falsely identify 2 halos with core-collapse times exceeding 20 Gyr as core-collapsing. One of these two halos is not predicted to collapse for over $10^3$ Gyr, indicating that the growth in central density must be caused by some mechanism other than core-collapse. By tracing this halo's merger history back in time we found that this halo had a late-stage merger that coincided with a significant increase in central density. These results show that central density, as a marker of core-collapse, is not robust to halo growth through mergers. Additionally,  we find that this method fails to identify 4 halos found to be core-collapsing in Figure \ref{fig:time_coll}. We note that all of these halos are UFDs with halo masses $M_{halo} < 2\times10^8 M_\odot$, at these low masses the central density increases very little at the radii we measure.  We conclude that using the growth in central density as a metric to identify core-collapsing halos is less accurate than measuring the inner density slope, particularly for the lowest-mass halos in our sample.  

\section{Turndown in Central Density at Late Times} \label{sec:app_A}
\begin{figure*}

  \centering
  \includegraphics[width=1\linewidth]{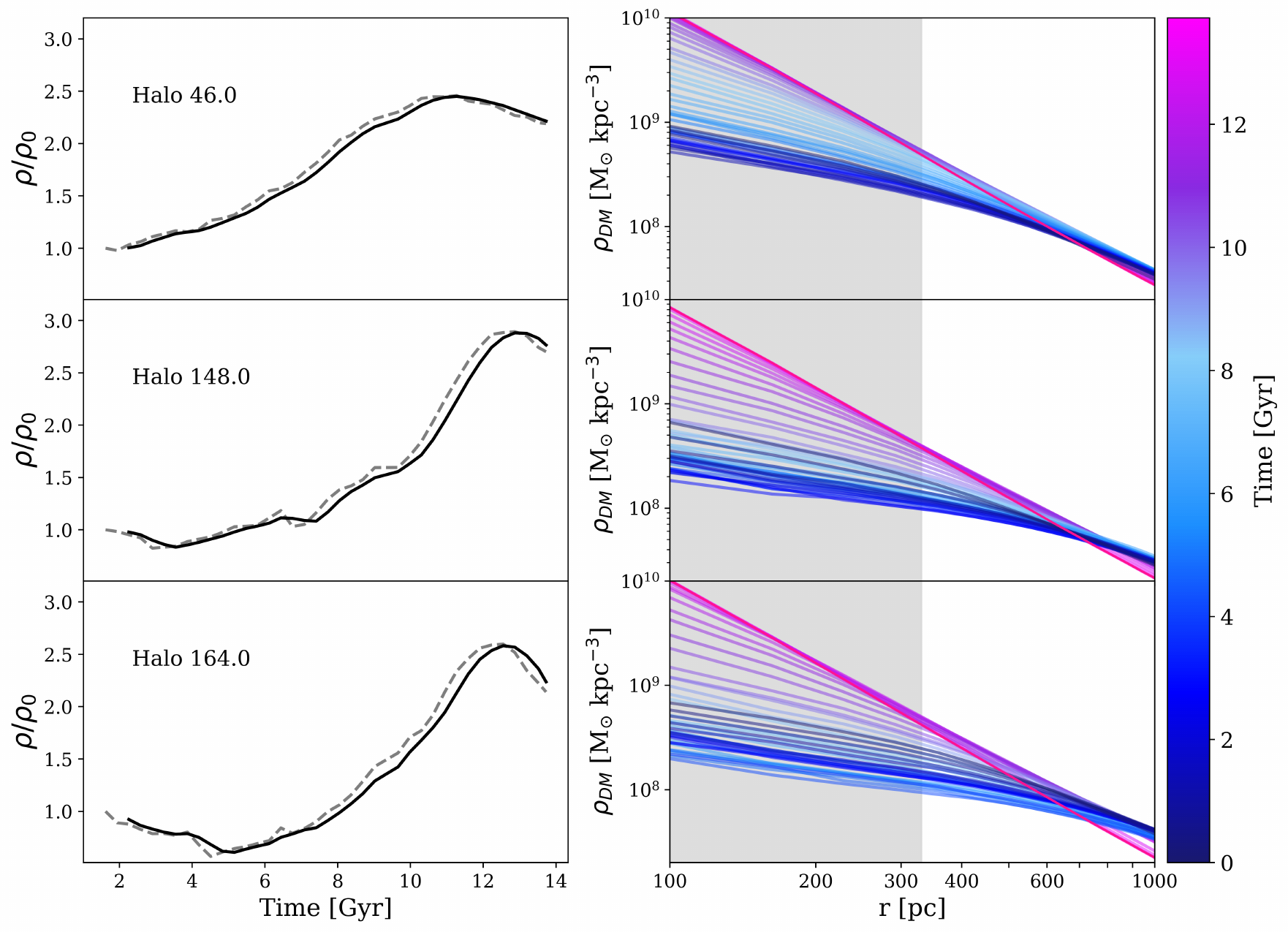}
  \caption{Left panel: Average central density within 0.35 kpc divided by initial average central density traced through time. Dashed black line is the raw profile while the solid line is smoothed. Right panel: Zoomed in DM density profiles plotted at different time-steps colored by time, with dark blue corresponding to early times (high z) and pink corresponding to late times (low z). The gray shaded region marks the region where the profile is unresolved, at radii less than 0.33 kpc. Halos 46, 148, and 164 are plotted in the top, middle, and bottom rows respectively.}
  \label{fig:turndown_46}
\end{figure*}

\indent Figure \ref{fig:turndown_46} shows the evolution of the DM density profiles for halos 46, 146 and 164. These are the three core-collapsed halos that experience a turndown in central density at late times. In the left column of the figure we plot the central density traced through time where the central density is divided by initial central density and calculated using the same methodology as in Figure \ref{fig:trace_core}. The right column tracks the DM density profile of each halo through time, zooming in on the inner regions of the halo. The density profile is smoothed by calculating density from the mass enclosed profile. The gray shaded area extends out to a radius of 0.33 kpc and marks where the profile is considered unresolved. The panels adopt the same color bar with pink corresponding to late times and dark blue corresponding to early times. 
\newline \indent In the top left panel of Figure \ref{fig:turndown_46} we show that during the initial phases of core-collapse, the central density of halo 46 increases rapidly before reaching a plateau at around 10 Gyr. Shortly after, we show that the central density begins to decrease. The top right panel of Figure \ref{fig:turndown_46} shows that this unexpected turndown in DM density results from the process of gravothermal core-collapse. As the inner DM halo grows denser and lower in mass, particles get kicked out of the halo center in order for it to contract. Because we measure central density at a fixed radius, as the halo contracts the central core shrinks to be inside of our radius of measurement. This effect  could be improved with higher mass resolution of DM particles. At the innermost regions of the halo profile the central density continues increasing up to redshift zero. However, central density is defined in this work as the averaged density of the DM halo within 0.33 kpc of the halo center. At this radius, the density of the DM halo begins to decrease at late times in halos that have entered late stages of core-collapse. We suggest that by increasing the mass resolution of our SIDM particles, we could more confidently resolve the inner regions of our DM halos and measure central density at smaller radii. 
\newline \indent Similarly, we conclude from the middle and bottom rows of Figure  \ref{fig:turndown_46} that the turndown in central densities of halos 146 and 164 results from the same phenomenon demonstrated previously in halo 46. 
 
\section{Resolution dependence on particle number}
 \label{sec:app_B}
\begin{figure*}
\centering
\includegraphics[width=1\linewidth]{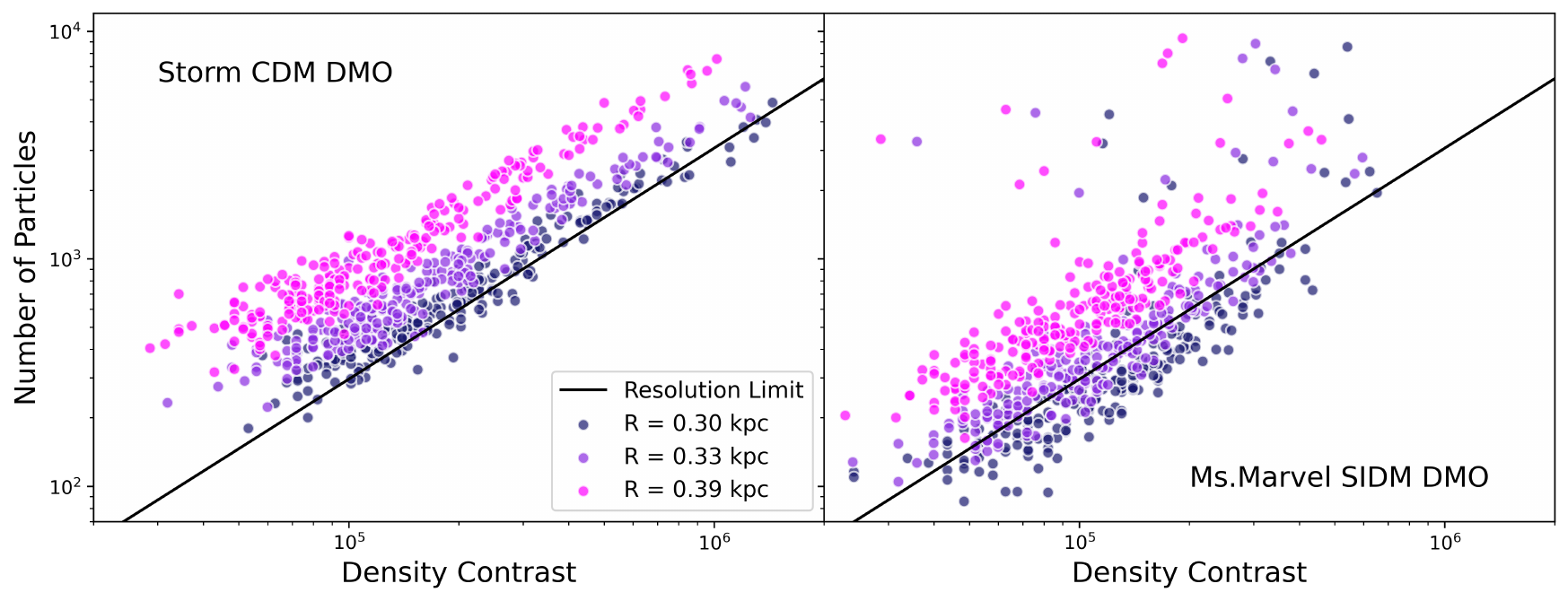}
\caption{Left side: Number of DM particles enclosed within a radius of R (kpc) plotted against density contrast measured at that radius for Storm CDM DMO halos. Dark blue, purple, and pink points represent the number of particles enclosed within 0.30, 0.33, and 0.39 kpc respectively. The black line indicates the quantity of particles necessary to resolve a CDM halo for a given density contrast as described by \cite{ShrinkSphere}. We note that \citep{ShrinkSphere} implement a CDM-based resolution criteria which may have limited applicability in SIDM simulations. 
Right side: Number of DM particles plotted against density contrast for Ms.Marvel SIDM DMO halos.}
\label{fig:res}
\end{figure*}
\indent Simulations face various challenges accurately resolving the centers of $\Lambda$CDM halos within one to a few times the softening length. For example, dynamical heating of DM particles at the centers of halos causes slopes of central DM density profiles to become artificially shallow at these radii. However, \cite{ShrinkSphere} demonstrate that for CDM simulations it is important to also consider the number of DM particles enclosed within a given radius and at a given density contrast. They found that CDM halos with too few particles at their centers have artificially low central densities. These artificially low central densities could result in shallower central inner density slopes, making halos appear cored when they are not. The following section tests the limits of spatial resolution in our simulations and justifies our adoption of $r = 0.33$ kpc as the smallest central radius that remains resolved for the CDM halos in our sample. 
\newline\indent All Storm CDM DMO halos in our sample are well resolved within 0.33 kpc. We find that, when applying the same resolution criteria trained on CDM simulations to SIDM halos, the full sample of Ms.Marvel DMO halos only become resolved at radii equal to or greater than 0.39 kpc. We identify resolved halos using criteria from \cite{ShrinkSphere} in Figure \ref{fig:res}, which plots the number of DM particles against the density contrast at a given radius for each Storm CDM DMO or Ms.Marvel DMO halo. The density contrast $(\frac{\rho(r)}{\rho_{crit}})$ is found by dividing the DM density by the critical density of the universe, while the number of DM particles enclosed within this given radius is plotted along the y-axis. Dark blue, purple, and pink points represent measurements for each halo within 0.30 kpc, 0.33 kpc, and 0.39 kpc, respectively. The black line marks the minimum number of particles required for a halo to become fully resolved for a given density contrast according to the criteria of \cite{ShrinkSphere}. Resolved halos lie above this threshold and unresolved halos lie below. We derive this threshold by plotting a straight line in log-log space between 100 particles with $\frac{\rho(r)}{\rho_{crit}} = 10^{4.5}$ and 3000 particles with $\frac{\rho(r)}{\rho_{crit}} = 10^6$ \citep{ShrinkSphere}. Using this criterion, we find that only 39 of our 235 Storm CDM DMO halos remain unresolved within 0.30 kpc. We note that for Ms.Marvel DMO halos, the relationship between the number of DM particles and the density contrast is more scattered relative to their CDM counterparts, and a larger number of halos are unresolved down to smaller radii. Despite these key differences, the majority of Ms.Marvel DMO halos are also resolved within 0.33 kpc. 
\newline\indent The resolution criteria published by \cite{ShrinkSphere} are derived using CDM simulations, meaning that they do not consider how SIDM self-scattering may affect the number of DM particles or the density within the centers of resolved halos. Without extensive resolution tests, it is difficult to anticipate whether the same resolution criteria used in CDM simulations apply to Ms.Marvel DMO halos. Because our entire sample of Storm CDM DMO halos is fully resolved down to 0.33 kpc, we chose to adopt a universal spatial resolution limit of 0.33 kpc for their Ms.Marvel DMO counterparts. Furthermore, SIDM halos in the advanced stages of core-collapse face additional resolution challenges \citep{zeng2025tillcorecollapsesevolution} and in these cases the central density reached by the halo should be understood as a lower limit. 
\newline\indent  Previous literature has already shown that modeling SIDM halos in the advanced stages of core-collapse poses many challenges \citep{YangYu2022,Zhong2023,Mace2024,Palubski2024,Fischer_2024,fischer2025}. Some works have already introduced recommendations for DM particle mass resolution \citep{Mace2024,Palubski2024}, gravitational softening lengths \citep{fischer2025}, and time-stepping \citep{fischer2025} which reduce errors in energy conservation and core-collapse time. Future work extensively testing the resolution limits of simulated SIDM halos will improve the implementation of SIDM models in simulations and reveal how simulation prescriptions impact the behavior of SIDM halos. 

\section{Density Profiles of Ms.Marvel Halos}
\begin{figure*}
\centering
\includegraphics[width=1\linewidth]{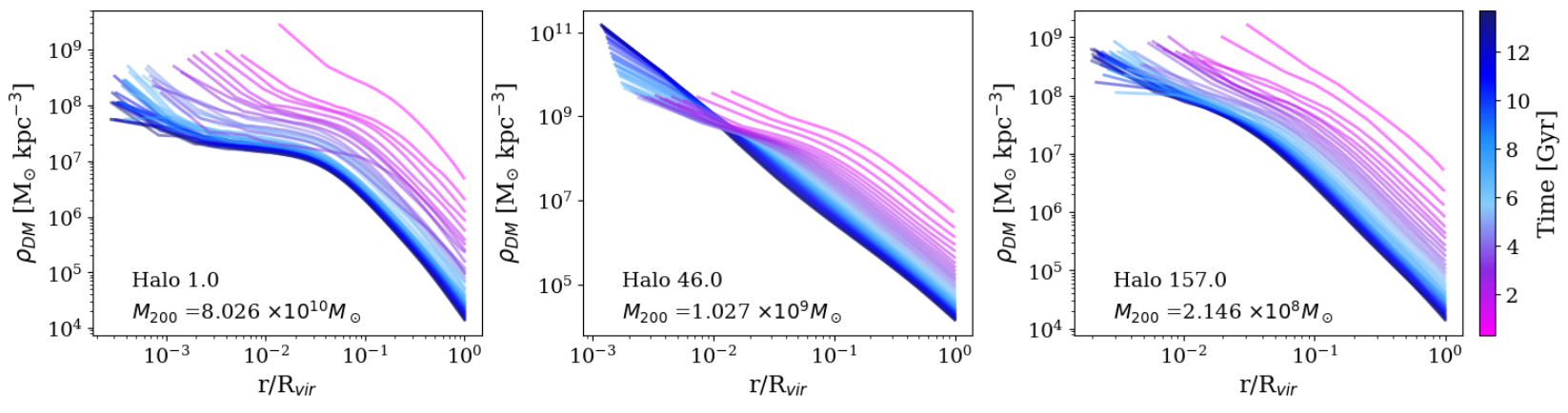}
\caption{Left panel: the DM density profiles of Halo 1 from the Ms.Marvel DMO simulation (the most massive halo in our sample) traced through time. The radius is normalized by the virial radius at each timestep and the evolution over time is illustrated by color, with pink profiles corresponding to earlier times and darker blue profiles corresponding to late times. Middle panel: the DM density profile of the most massive halo that we identified as undergoing core-collapse in our sample. Right panel: a small halo that follows the average evolution track of core-forming SIDM halos in this mass bin.}
\label{fig:density_prof}
\end{figure*}

\indent In Figure \ref{fig:density_prof} we show three density profiles from the Ms.Marvel DMO simulation to illustrate how the density profiles of SIDM halos evolve over time across different mass ranges. Halo 1 is the most massive halo in our sample of Ms.Marvel halos and is core-forming at redshift zero. We see that in the left-most panel this halo forms its core at early times and has a lowered central density. Halo 46 is a medium-mass halo in our sample, but is the highest-mass halo that we identify as having entered the core-collapse phase. We see in the middle panel that this halo begins forming a core before initializing core-collapse and becoming extremely centrally concentrated at late times. In the right-most panel we show the evolution of halo 157, a low-mass halo from our sample that is core-forming at late times. We see that the central core of this halo remains much smaller and less pronounced than more massive halos, however, it is much less centrally dense and cuspy than the core-collapsed halos.

\acknowledgments
A.C.E. is supported by the National Aeronautics and Space Administration Future Investigators in NASA Earth and Space Science and Technology (FINESST) (22-ASTRO22-0096). FDM and BV acknowledge support from NSF grant PHY2013909.
FDM is grateful for the hospitality of Perimeter Institute where part of this work was carried out. Research at Perimeter Institute is supported in part by the Government of Canada through the Department of Innovation, Science and Economic Development and by the Province of Ontario through the Ministry of Colleges and Universities. This research was also supported in part by the Simons Foundation through the Simons Foundation Emmy Noether Fellows Program at Perimeter Institute. This research was supported in part by grant NSF PHY-2309135 to the Kavli Institute for Theoretical Physics (KITP).
AMB acknowledges support from NSF grant AST-2306340, and by grant FI-CCA-Research-00011826 from the Simons Foundation.  AHGP acknowledges support from NSF grants  AST-2008110 and AST-2510899.  
Some of the simulations were performed using resources made available by the Flatiron Institute. The Flatiron Institute is a division of the Simons Foundation. 
This work used Stampede2 at the Texas Advanced Computing Center (TACC) through allocation MCA94P0181642 from the Advanced Cyberinfrastructure Coordination Ecosystem: Services \& Support (ACCESS) program, which is supported by U.S. National Science Foundation grants 2138259, 2138286, 2138307, 2137603, and
2138296.

 \bibliographystyle{JHEP}
\bibliography{bibliography.bib}

\end{document}